\documentstyle[epsfig, subfigure, lscape, onecolumn]{mn}
\newcommand{\be}{\begin{equation}}
\newcommand{\ee}{\end{equation}}
\def\la{\buildrel < \over {_{\sim}}}
\def\ga{\buildrel > \over {_{\sim}}}

\title[]{Probing the origin of giant radio halos through radio 
and $\gamma$-ray data : the case of the Coma cluster}

\author[G. Brunetti, et al.]{
G. Brunetti, $^{1}$\thanks{E-mail: brunetti@ira.inaf.it}
P. Blasi, $^{2}$\thanks{E-mail: blasi@arcetri.astro.it}
O. Reimer, $^{3}$\thanks{E-mail: Olaf.Reimer@uibk.ac.at}
L. Rudnick, $^{4}$\thanks{E-mail: larry@astro.umn.edu}
A. Bonafede, $^{5}$\thanks{E-mail: a.bonafede@jacobs-university.de}
S. Brown$^{6}$\thanks{Bolton Fellow, E-mail: Shea.Brown@csiro.au}\\
{$^1$ \it INAF-IRA, Via Gobetti 101, I-40129 Bologna, Italy}\\
{$^2$ \it INAF-Osservatorio Astrofisico di Arcetri, Largo E. Fermi, 
5, 50125 Firenze, Italy}\\
{$^3$ \it Institut f\"ur Astro- und Teilchenphysik,
Leopold-Franzens-Universit\"at Innsbruck, A-6020 Innsbruck, Austria}\\
{$^{4}$ \it Minnesota Inst. for Astrophysics, School of Physics \&
Astronomy, Univ. of Minnesota, 116 Church Street SE, Minneapolis, MN
55455, USA}\\
{$^5$ \it Jacobs University Bremen, Campus Ring 1, D-28759 Bremen, 
Germany}\\
{$^{6}$ \it CSIRO Astronomy \& Space Science, P.O. Box 76, Epping NSW 1710,
Australia}\\ 
}
\begin{document}
\date{Accepted -----. Received -----}
\maketitle

\label{firstpage}

\begin{abstract}
We combine for the first time all available information about the 
spectral shape and morphology of the radio halo of the Coma cluster 
with the recent $\gamma$-ray upper limits obtained by 
the Fermi-LAT and with the magnetic field 
strength derived from Faraday rotation measures. 
We explore the possibility that the radio emission is due to synchrotron 
emission of secondary electrons.
First we investigate the case of pure secondary models
that are merely based on the mechanism of continuous injection of secondary
electrons via proton-proton collisions in the intra-cluster medium.
We use the observed spatial distribution of the halo's radio 
brightness to constrain the amount of 
cosmic ray protons and their spatial distribution in the cluster that are
required by the model.
Under the canonical assumption that the spectrum of cosmic rays
is a power-law in momentum and that the spectrum of secondaries 
is stationary,
we find that the combination of the steep spectrum of cosmic ray protons
necessary to explain the spectrum of the halo and the very
broad spatial distribution (and large energy density) of 
cosmic rays result
in a $\gamma$--ray emission in excess of present limits, unless the
cluster magnetic field is relatively large. 
However this large magnetic field required to not violate present 
$\gamma$--ray limits appears inconsistent with that derived from 
recent Faraday rotation measures. 
Second we investigate more complex models in which the cosmic rays 
confined diffusively in the Coma cluster and their secondary electrons
are all reaccelerated by MHD turbulence.
We show that under these conditions it is possible to explain the radio 
spectrum and morphology of the radio halo and to predict $\gamma$-ray fluxes 
in agreement with the Fermi-LAT upper limits without tension with 
present constraints on the cluster magnetic field.
Reacceleration of secondary particles 
also requires a very broad cosmic ray spatial profile, much flatter than 
that of the intracluster medium, at least provided that both the turbulent
and magnetic field energy densities scale with that of the intracluster 
medium. However, this requirement can be easily alleviated if we assume that 
a small amount of (additional) seed primary electrons are reaccelerated 
in the cluster's external regions, or if we adopt flatter scalings of the 
turbulent and magnetic field energy densities with distance from the 
cluster center.
\end{abstract}


\section{Introduction}

Galaxy clusters host several potential accelerators of cosmic ray (CR) 
electrons and protons, from ordinary galaxies to active galaxies (AGN) 
and cosmological shock waves, driven in the intracluster medium (ICM) 
during the process of hierarchical cluster formation (see Blasi et al. 
2007 for a review).

The long lifetime of CR protons (or nuclei) in the ICM and the large 
geometrical size of the magnetized region of galaxy clusters make them 
efficient storage rooms for the hadronic component of CRs produced within 
their volume (V\"olk et al. 1996, Berezinsky et al. 1997, 
Ensslin et al. 1997). The accumulation of CRs inside clusters over 
cosmological times leads to the assumption 
that an appreciable amount of energy 
may be stored in the ICM in the form of non-thermal particles. 
If this energy is sufficiently high, the flux of $\gamma$ radiation induced 
by the production and decay of neutral pions may reach potentially 
detectable levels, thereby providing us with a powerful diagnostic tool of 
the CR energy content of clusters (Colafrancesco \& Blasi 1998, 
Blasi \& Colafrancesco 1999, V\"olk \& Atoyan 1999, Miniati 2003, 
Pfrommer \& En\ss lin 2004, Wolfe et al. 2008).

So far only upper limits to the $\gamma$-ray emission from galaxy clusters 
have been obtained (Reimer et al.~2003; Perkins et al. 2006, 
Aharonian et al. 2009a,b; Aleksic et al. 2010; Ackermann et al. 2010)
\footnote{See however Han et al. 2011 for Virgo}.
These upper limits, together with several constraints from complementary 
approaches based on radio observations lead us to conclude that 
CR protons contribute less than a few percent
of the energy of the ICM, at least in the central Mpc--size region
(Reimer et al. 2004, Brunetti et al. 2007, 2008, Brown et al. 2011,
Aleksic et al. 2011).

CR electrons are very well traced in the ICM through their radio emission 
which appears in the form of diffuse (Mpc scale) 
synchrotron {\it giant radio halos} from the cluster X-ray emitting regions, 
and {\it relics}, typically in the clusters' peripheral regions 
(see Ferrari et al. 2008, Venturi 2011, for recent reviews on observations).

\noindent
Giant radio halos are the most spectacular and best studied non-thermal
large scale phenomena in the universe. They appear in about $1/3$ of the 
most massive galaxy clusters (e.g. Giovannini et al. 1999, 
Kempner \& Sarazin 2001, Cassano et al. 2008), in a rather clear connection 
with dynamically disturbed systems, while ``off-state'' clusters 
(those with no evidence of diffuse emission) are generally more 
relaxed (Cassano et al. 2010a and references therein). 
The connection between cluster mergers and radio halos suggests that 
such emission traces the hierarchical cluster assembly and probes the 
dissipation of gravitational energy during the dark matter-driven 
mergers that lead to the formation of clusters. 
The physical mechanisms responsible for the generation and evolution of 
radio halos are still a matter of debate, but two main
lines of thought have been 
developed throughout the years.

\noindent
One is based on the idea that seed electrons 
may be re-accelerated by turbulence produced during merger 
events (Brunetti et al. 2001, Petrosian 2001). 
In this class of models the $\gamma$-ray emission is predicted to
be rather low, 
though it may be substantial under the hypothesis that the electron seeds
are secondaries produced in inelastic collisions of a subdominant 
hadronic CR component in the ICM (Brunetti \& Blasi 2005; 
Brunetti \& Lazarian 2011). 

\noindent
The second line of thought is based on the idea of clusters as storage 
rooms of CR protons: the radio halos may be generated as a result of 
synchrotron emission of secondary electrons and positrons from pp 
collisions (Dennison 1980,
Blasi \& Colafrancesco 1999, Pfrommer \& En\ss lin 2004). 
On one hand this idea serves as a solution to the problem of the large 
spatial dimensions of the radio emitting region, larger than the typical 
loss length of electrons: secondary electrons and positrons are 
produced {\it in situ} in inelastic CR collisions and radiation 
is produced near the production region. 
On the other hand, secondary models do not explain in a {\it natural} way 
the observed association between cluster mergers and giant 
radio halos, in that CRs accumulate inside the ICM on cosmological 
time scales and not in direct connection with acceleration events 
such as those associated with mergers\footnote{see however Ensslin et al.
2011, where diffusion/transport of CRs is studied under peculiar conditions}. 
One proposal is that 
during merger events the cluster magnetic field becomes larger than 
in the quiescent state, thereby turning the halo on (Kushnir et al. 2009, 
Keshet \& Loeb 2010), although studies based on Rotation Measures
of clusters' and background radio sources disfavour this scenario
(see Bonafede et al.~2011a and ref. therein).
In fact, some pieces of observations put tension on a {\it pure} hadronic 
origin of radio halos. These include the steepening in the spectrum 
(or the very steep spectrum) of several halos (Schlickeiser et al. 1987,
Thierbach et al. 2003, Reimer et al. 2004, 
Brunetti et al. 2008, Dallacasa et al. 2009,
Giovannini et al. 2009, Macario et al. 2010, van Veeren et al. 2011) and the 
very large spatial extent of several halos (or their flat radio-brightness 
distribution) (Brunetti 2004,
Murgia et al. 2009, Donnert et al. 2010a, Brown \& Rudnick 2011); 
in all cases observations would imply that the energy budget of CR protons 
is uncomfortably large, at least assuming that clusters are magnetised 
at $\sim \mu$G level, consistent with the results of 
RM (see Bonafede et al.~2010 and references therein).

\noindent
The most distinct prediction of models of radio halos that are based on
secondary particles is the 
production of $\gamma$--rays (e.g., Blasi \& Colafrancesco 1999, 
Sarazin 2004, Pfrommer \& En\ss lin 2004, Brunetti 2009,
Brunetti \& Lazarian 2011, En\ss lin et al. 2011). 
Nowadays there is agreement on the fact that the abundance of secondaries
required to fit the spectrum of (at least some) radio halos should produce
$\gamma$--ray emission detectable with the sensitivity of the Fermi-LAT,
assuming low/medium level of cluster magnetic field (eg. Marchegiani
et al 2007, Pfrommer 2008, Brunetti 2009, Jeltema \& Profumo 2011).
Thus 
the combination of the available information on the spectrum and morphology 
of radio halos in conjunction with 
the limits on their $\gamma$-ray emission provides 
a powerful tool to gain insights into the origin of the halo emission.

\noindent
Following this pathway, in 
this paper we concentrate on the case of the Coma cluster, which 
represents a prototypical example of giant radio halos, with a wealth 
of data available on its spectrum and morphology. 
The incoming upper limits on the Coma $\gamma$-ray emission with the 
Fermi-LAT telescope are invaluable in imposing stringent limits on 
the amount of cosmic rays that can be stored in the intracluster medium 
and serve as sources of secondary electrons and positrons. 
In particular in this paper we combine for the first time 
all available information about the
spectral shape and morphology of the radio halo of the Coma cluster
with the recent $\gamma$-ray upper limits and magnetic field
strengths derived from Faraday rotation measures.
We show that the requirement of reproducing
the properties of the Coma radio halo in the context of a pure
secondary electron model leads to a large CR energetics and fluxes 
of $\gamma$-rays which results in a tension with the existing upper limits 
from the Fermi-LAT (Ackermann et al.~2010).
This situation is readily alleviated in the case of a
large cluster magnetic field,
however the magnetic fields required by the model are appreciably larger 
than those inferred from recent Faraday rotation measures. 

\noindent
This tension disappears when including the effect of turbulent
reacceleration in combination with the process of injection of
secondary particles.
We adopt a physically motivated 
picture in which secondary products of cosmic ray interactions are
reaccelerated by MHD turbulence during clusters mergers.
In this sense we value the physical insight behind the concept of cosmic 
ray confinement and the production of secondary electrons, but we do not 
assume that these particles are ``directly'' responsible for the formation of 
the radio halo. 
We find that even reacceleration models of this type require a broad 
spatial distribution of the parent cosmic rays, but the expected 
$\gamma$-ray 
fluxes are well consistent with the Fermi-LAT upper limits.

The paper is organized as follows: we discuss the hadronic model of 
radio emission in \S \ref{sec:hadro} and the reacceleration model 
in \S \ref{sec:reacc}. A critical discussion of our results 
is provided in \S \ref{sec:disc}.
A $\Lambda$CDM
cosmology ($H_{o}=70\,\rm km\,\rm s^{-1}\,\rm Mpc^{-1}$,
$\Omega_{m}=0.3$, $\Omega_{\Lambda}=0.7$) is adopted throughout the paper.

\section{Pure hadronic models}
\label{sec:hadro}

\subsection{Formalism : radio and $\gamma$--ray emission}

The decay chain responsible for the injection of secondary particles 
in the ICM due to p-p collisions is (e.g., Blasi \& Colafrancesco 1999):

$$p+p \to \pi^0 + \pi^+ + \pi^- + \rm{anything}$$
$$\pi^0 \to \gamma \gamma$$
$$\pi^\pm \to \mu^{\pm} + \nu_\mu ~~,~~ \mu^\pm\to e^\pm \nu_\mu \nu_e \, .$$

The initial inelastic scattering reaction occurs at threshold, 
namely it requires protons with kinetic energy larger 
than $T_p \approx 300$ MeV. The injection rate of pions is 
\begin{equation}
Q_{\pi}^{\pm,o}(E_{\pi^{\pm,o}},t)= n_{th} c 
\int_{p_{*}} dp N_p(p,t) \beta_p {{ F_{\pi}(E_{\pi},E_p) 
\sigma^{\pm,o}(p)}\over
{\sqrt{1 + (m_pc/p)^2} }},
\label{q_pi}
\end{equation}
where $n_{th}$ is the number density of thermal (target) protons,
$N_p$ is the spectrum of cosmic ray protons, 
and $F_{\pi}$ is the spectrum of pions from the collision between 
a CR proton of energy $E_p$ and thermal protons 
(taken from Brunetti \& Blasi 2005). 
The inclusive cross--section, $\sigma(p)$, is taken from the fitting 
formulae of Dermer (1986b), which allows us to describe separately 
the rates of generation of $\pi^{-}$, $\pi^{+}$, and $\pi^{o}$, 
and $p_* = \max \{ p_{_{tr} },\, p_{\pi} \}$, where $p_{tr}$ is the 
threshold momentum of protons for pion production (different for 
charged and neutral pions). 

Charged pions decay into muons and then
secondary pairs (electrons and positrons). 
If secondaries are not accelerated by other mechanisms and the physical 
conditions in the ICM do not change on time-scales shorter than 
the electrons' radiative time \footnote{see Keshet (2010) for a discussion 
on effects due to variations of magnetic fields on shorter time-scales}, 
their spectrum approaches a stationary distribution because of the 
competition between injection and energy losses (e.g.,
Dolag \& Ensslin 2000), and one can write:

\begin{equation}
N_e^{\pm}(p)=
{1 \over
{\Big| \left( {{dp}\over{dt}} \right)_{\rm loss} \Big| }}
\int_{p}^{p_{\rm max}}
Q_e^{\pm}(p) dp \, ,
\label{sec_stat}
\end{equation}

\noindent
where $Q_e^{\pm}$ is the injection rate of secondaries 
(Blasi \& Colafrancesco 1999; Moskalenko \& Strong 1998), and 
radiative losses, which are dominant for $\gamma > 10^3$ electrons 
in the ICM, can be written as (Sarazin 1999):

\begin{equation}
\Big| \left( {{ d p }\over{d t}}\right)_{\rm loss} \Big|
\simeq 3.3 \times 10^{-32}
\big( {{p/m_e c}\over{300}} \big)^2 \left[ \left( {{ B_{\mu G} }\over{
3.2}} \right)^2 + (1+z)^4 \right] \, .
\label{loss}
\end{equation}

Assuming a power law (momentum) distribution of CR protons, 
$N_p(p) = K_p p^{-s}$, the spectrum of secondaries 
at high energies, $\gamma > 10^3$, 
is $N_e(p) \propto p^{-(s+1)} {\cal F}(p)$, 
where ${\cal F}$ accounts for the log--scaling of the p-p cross--section 
at high energies and causes the spectral shape to be slightly flatter 
than $p^{-(s+1)}$ (Brunetti \& Blasi 2005). 
The synchrotron spectrum from secondary e$^{\pm}$ 
is (Rybicky \& Lightman 1979):

\begin{equation}
J_{syn}(\nu) = \sqrt{3} {{e^3}\over{m_e c^2}} B
\int_0^{\pi/2} d\theta sin^2\theta \int dp N_e(p)
F\big( {{\nu}\over{\nu_c}} \big)
\simeq C_{syn}(\alpha, T) X n_{th}^2
{{B^{1+\alpha} }\over{B^2 + B_{cmb}^2}} \nu^{-\alpha} \, ,
\label{jsyn}
\end{equation}
where $C_{syn}$ is a constant, $X=\epsilon_{CR}/\epsilon_{ICM}$ 
is the ratio of the energy densities of CR protons to thermal protons, 
$F$ is the synchrotron kernel, $\nu_c$ is the critical frequency, 
and $\alpha \simeq s/2 -\Delta$ is the synchrotron spectral index, 
where $\Delta \sim 0.1$ (in the conditions
of interest for our paper) is due to 
the log-scaling of the cross--section.

The spectrum of $\gamma$-rays produced by secondary particles is dominated 
by the decay of the secondary $\pi^0$ (Dermer 1986ab; 
Blasi \& Colafrancesco 1999, Blasi 2001) :

\begin{equation}
Q_{\gamma}(E_{\gamma})
= 2 \int_{E_{min}}^{E_p^{max}}
{{ Q_{\pi^0}(E_{\pi^0}) }\over{
\sqrt{E_{\pi}^2 - m_{\pi}^2 c^4} }}
d E_{\pi} \, ,
\label{qgamma}
\end{equation}

\noindent
where $E_{min} = E_{\gamma} + m_{\pi}^2 c^4 / (4 E_{\gamma})$. 
Under the same assumptions as Eq.\ref{jsyn}, the $\gamma$-ray emissivity 
for $h\nu_{\gamma} >$ GeV is:

\begin{equation}
J_{\gamma}(\nu) 
\simeq C_{\gamma}(s, T) X n_{th}^2 \nu^{-\tilde{s}} \, ,
\label{jgamma}
\end{equation}

\noindent
where $C_{\gamma}$ is a constant and $\tilde{s} \sim s -1$. 
This gives a minimum estimate of the cluster $\gamma$--ray emission 
since IC and non-thermal bremsstrahlung emission from secondary and 
primary electrons provide additional 
contributions (e.g. Sarazin 1999, Blasi 2001, Miniati 2003).

\subsection{Testing hadronic models for radio halos}

The ratio of the synchrotron and $\gamma$--ray cluster luminosities depends 
on the magnetic field in the emitting volume (from Eqs.~\ref{jsyn} 
and \ref{jgamma}). By considering reasonable assumptions on the magnetic 
field in galaxy clusters several authors suggested that observations
by the Fermi-LAT 
may have the chance to detect $\gamma$-rays from nearby clusters hosting 
radio halos (Pfrommer \& En\ss lin 2004, Marchegiani et al. 2007,
Pfrommer 2008, Wolfe et al. 2008, Brunetti 2009).

More recently, Jeltema \& Profumo (2011) analysed the impact of 
limits on the $\gamma$-ray emission from clusters of galaxies (Ackermann
et al.~2010) on 
hadronic models for the origin of cluster radio halos, deriving lower 
limits on the average cluster magnetic field. In some cases of nearby 
radio halos they find that these lower limits are close to (or larger than) 
the magnetic field values inferred from Faraday rotation measures, 
thereby placing tension on the hadronic origin of radio halos. 
In their analysis 
Jeltema \& Profumo used the total radio luminosity and the
$\gamma$-ray upper limits.

Here we discuss the possibility that even more stringent limits may 
be obtained by using the information about the morphology of the 
radio emission {\it in addition} to the total luminosity. 
Giant radio halos have flat brightness distributions, in several 
cases flatter than the thermal X-ray emission of the hosting clusters
(e.g. Govoni et al. 2001, Feretti et al. 2001, Brown \& Rudnick 2011), 
implying that most of the synchrotron emission is produced in the external 
regions where the magnetic field is smaller.
Thus, we anticipate a larger energy budget of CR protons 
compared with that needed to explain only the total luminosity 
of radio halos (Brunetti 2004, Donnert et al. 2010a,b). 
The consequence is also a larger $\gamma$-ray luminosity from the hosting 
clusters (Donnert et al. 2010a).

\noindent
Given these premises, in Sect.~2.3 we attempt to analyse the impact 
of the Fermi-LAT upper limits from the Coma cluster. We summarize our 
assumptions as follows:

\begin{itemize}

\item[{\it i)}]
The spectrum of cosmic ray protons is assumed to be a power 
law in momentum, as results from most acceleration mechanisms, 
$N(p) = K_p p^{-s}$;

\item[{\it ii)}]
The spatial distribution of the gas in the ICM is taken to follow 
the {\it observed} $\beta$--model that is parameterised according
to thermal density at the cluster center, $n_{th}(0)$, the 
core radius, $r_c$ kpc, and $\beta$;

\item[{\it iii)}]
The spatial distribution of the energy density of CR protons in the 
cluster volume is parametrized 
as $\epsilon_{CR} \propto \epsilon_{ICM}^{1+f}$;

\item[{\it iv)}]
The magnetic field strength is assumed to scale with location $r$ 
in the ICM according with 
$B(r)=B_0 \left[\epsilon_{ICM}(r)/ \epsilon_{ICM} (0)\right]^{\eta}$,
$\eta=0.5$ implies that the magnetic field energy density scales with 
the thermal one.

\end{itemize}

\noindent
Within these assumptions the synchrotron emissivity from secondary 
particles (from Eq.~\ref{jsyn}) is:

\begin{equation}
J_{syn}(\nu) \propto
n_{th}^{2+f} T^{1+f}
{{ ( n_{th} T)^{\eta(1+\alpha)}}\over{
C_B^2 ( n_{th} T)^{2 \eta} + B_{cmb}^2 }},
\label{JsynS}
\end{equation}
which implies a synchrotron brightness (assuming $T$ constant 
on $\sim 1$ Mpc scale to derive simple scalings) 
$I_{syn} \propto (1 + x^2)^{-3 \tilde{\beta} +1/2}$, 
where $x$ is the projected distance in units of the core radius, 
$C_B$ is a constant ($B = C_B ( n_{th} T)^{\eta}$), 
and $\tilde{\beta} = \beta (2 +f +(\alpha -1) \eta)/2$ 
and $\tilde{\beta} =\beta (2 +f +(\alpha +1) \eta)/2$ in the 
case $B^2(x) \gg B_{cmb}^2$ and $B^2(x) \ll B_{cmb}^2$, respectively.

For a given model of magnetic field distribution, $(B_0,\eta)$, 
the three key parameters characterizing the radio emissivity and then
the predictions for the $\gamma$-ray emission are thus :

\begin{itemize}

\item[1.]
$s$: the logaritmic slope of the momentum distribution of CR protons, 
determined
from measurements of the radio spectral index $\alpha$.
Typically $s \sim 2.3-3$ (for some halos with ultra-steep 
spectrum $s > 3$);

\item[2.]
$f$: the non-linear scaling of the energy density of CR protons 
with the thermal
ICM energy density, derived from the shape of the
synchrotron brightness profile of the radio halo;

\item[3.]
$K_p$: the normalization of the momentum distribution of CR protons,
determined from the radio luminosity of the halo.

\end{itemize}

\noindent
The outcome of this procedure is the expected $\gamma$-ray 
emission from $\pi^0$--decays.

\subsection{Application to the Coma radio halo}
\label{sec:morph}

\begin{figure}
\begin{center}
{
\includegraphics[width=0.4\textwidth]{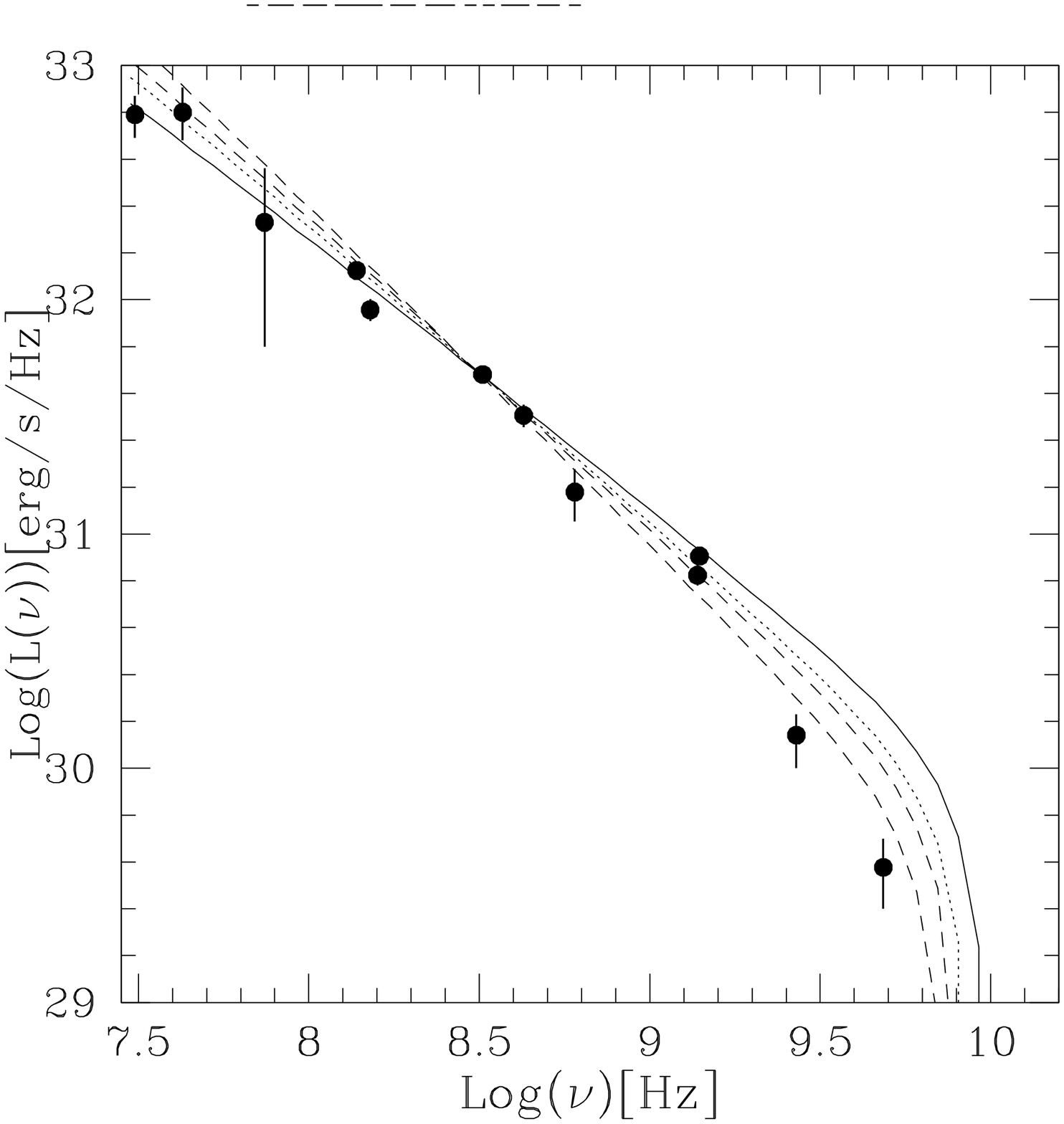}
\includegraphics[width=0.4\textwidth]{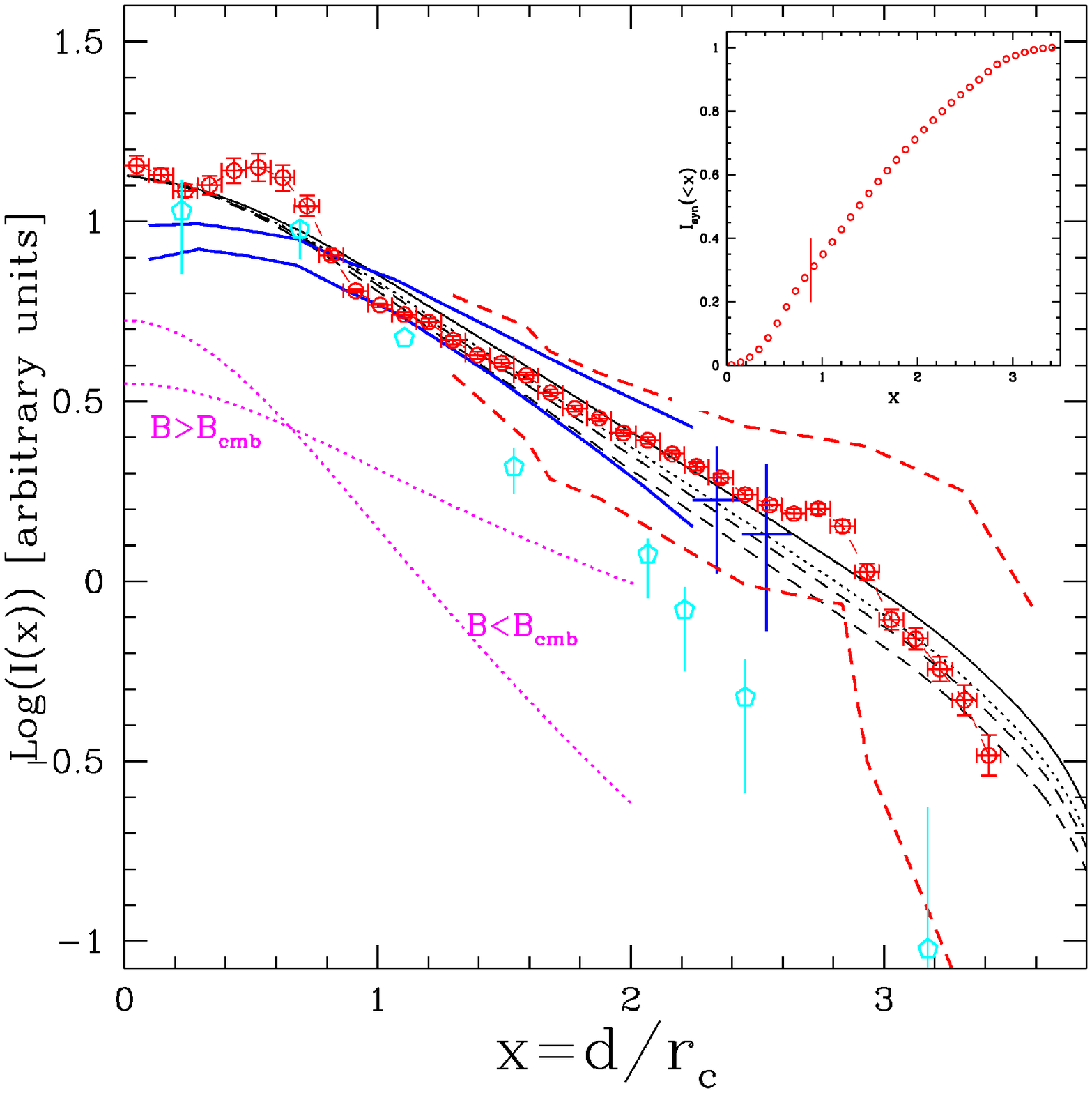}}
\end{center}
\caption{{\bf Left}: Spectrum of the Coma halo from hadronic models 
assuming a slope $s=$2.6 (solid), 2.85 (dotted), 3, 3.25 (dashed)
(compilation of data points taken from Pizzo 2010). 
Models are normalised at 350 MHz. The steepening at higher frequencies 
in the models is due to the thermal SZ--decrement (see text). 
{\bf Right}: The measured brightness (azimuthal averaged) profile of the 
Coma halo (in arbitrary units) at 350 MHz, using observations in
Brown \& Rudnick (2011), is shown as a function of
distance in units of core radius, $r_c = 270$ kpc (red points).
Red dashed lines mark the upper and lower envelope of the measured
profile taken along different directions (from Brown \& Rudnick).
The new measurements are compared with previous data at 350 MHz
from Govoni et al.(2001) (solid, blue and crosses) and 
with data at 1.4 GHz from Deiss et al.(1997) (cyan points).
The brightness profile from hadronic models is shown assuming 
$B_0 \sim 5 \mu$G, $\eta =0.5$ and $f=-1.6$ for $s=$2.6 
(same line--type as in the left panel). The inset shows the 
cumulative flux profile calculated using the Brown \& Rudnick azimuthal 
averaged profile, the vertical line shows the transition between 
the regime of synchrotron--dominance (smaller distances) 
and IC--dominance (larger distances from cluster center). 
Finally the (magenta) dotted--lines show the expected shape of the 
brightness profile assuming hadronic models with $f=-1.0$, with 
synchrotron--dominance and IC--dominance (scalings derived
in the paragraph below Eq.7); for display purposes the
magenta profiles are not normalised to the data.}
\label{fig:hadroComa}
\end{figure}

The radio halo in the Coma cluster (Coma C) is the prototypical cluster 
radio halo (e.g., Willson 1970, Giovannini et al. 1993). It is unique in 
that its spectrum is measured over a wide frequency 
range (Fig. \ref{fig:hadroComa}, left panel). The spectrum shows 
a significant steepening at high frequencies: a power-law that fits the 
data at lower frequencies overestimates the flux measured at 2.7 and 
5 GHz by a factor 2 and 4 respectively (Schlickeiser et al. 1987, 
Thierbach et al. 2003). In secondary electron models the electron 
spectrum extends in principle to very high energies, therefore no 
intrinsic cutoff is expected in the radio spectrum. 
The observed spectral steepening is therefore inconsistent with 
secondary electron models (Schlickeiser et al. 1987, Brunetti et al. 2001, 
Petrosian 2001, Blasi 2001, Reimer et al 2004) unless unnatural assumptions 
are made on the CR proton spectrum. 
En\ss lin (2002) suggested that the steepening at higher frequencies 
may be due to the thermal SZ-decrement (suppression of the CMB spectrum 
seen in the direction of the cluster from the up-scattering 
of the CMB photons due to Compton scattering with 
the thermal electrons in the ICM). Calculations carried out 
by Brunetti (2004) and Reimer et al. (2004) 
showed that this is not the case when the SZ estimate is limited 
to the region covered by the halo emission. 
More recent investigations, based 
on numerical simulations confirm these calculations showing 
that the effect of the SZ decrement on the Coma spectrum at 2.7 and 5 GHz 
is not sufficient to explain the observed 
spectral break (Donnert et al. 2010a).

The halo shows a fairly regular morphology on a  
scale $\sim$1.5 Mpc with a radio brightness profile $I_{syn}(r)$ that is 
flatter than the X-ray (thermal) profile $I_X$ 
(Govoni et al. 2001, Brown \& Rudnick 2011). 
The global point-to-point power-law correlation between the 0.3 GHz 
radio brightness and the X-ray brightness, $I_{syn} \propto I_X^{0.64}$, 
implies that most of the radio emission is produced outside the 
cluster core, $r_c \sim 270$ kpc. The brightness profile of radio
halos is an important observable quantity that provides 
constraints on the spatial distribution of CRs and
magnetic fields in the cluster (Brunetti 2004, Pfrommert \& En\ss lin 2004,
Colafrancesco et al. 2005, Donnert et al 2010a).
We measure the radial profile of the Coma halo using the new WSRT observations
at 350 MHz by Brown \& Rudnick (2011) and obtain unprecedented constraints 
on the brightness distribution of the halo up to 3-3.5 $r_c$ 
distance from the cluster center.
The radial (azimuthally averaged) brightness profile is 
shown in Fig. \ref{fig:hadroComa} (right
panel) and is compared with other profiles available in the literature,
Govoni et al. (2001) at 350 MHz, and Deiss et al (1997) at
1.4 GHz. The new brightness profile provides constraints that are
significantly better than those from previous observations and, most
important, they extend to larger spatial scales. The brightness profile 
is much flatter than the one found by Deiss et al., that has been widely 
adopted in the literature  
(Pfrommer \& En\ss lin 2004, Colafrancesco et al 2005, Donnert et al 2010a),
implying that the amount of energy in the form of 
CRs and magnetic field at distances $\sim 2-3.5 \, r_c$ from the
cluster center is much larger than previously thought.

\noindent
The inset in the right panel of  Fig. \ref{fig:hadroComa} shows the
cumulative flux profile of the Coma halo (based on Brown \& Rudnick 2011).
Assuming a central value of the magnetic field $B_0 \sim 5 \mu$G 
and $\eta=0.5$ (Bonafede et al. 2010), about 70-80\% of the observed 
radio luminosity is produced in regions where $B^2 \ll B_{cmb}^2$, 
while synchrotron dominance is expected only inside the cluster's core. 
In the right panel of Fig. \ref{fig:hadroComa} the (magenta) dotted-lines 
show the case of a hadronic model of the radio 
halo, with the shape of the expected synchrotron spectrum chosen 
to match the observed synchrotron spectral index, 
$\alpha \sim 1.3$ (Deiss et al. 1997), $\beta = 0.75$, and a flat spatial 
distribution of CR protons ($f=-1$). 
In the relevant case $B^2 \ll B_{cmb}^2$, the expected brightness 
is steeper than the observed one, implying that the energy density 
of CRs must increase with distance (at least up to 
about $r \sim 2.5-3 \, r_c$) in orderd to fit the observed radio
profile. This has consequences on the total 
energy budget of CRs in the external regions of the cluster and 
on the expected $\gamma$-ray emission.

\begin{figure}
\begin{center}
{
\includegraphics[width=0.79\textwidth]{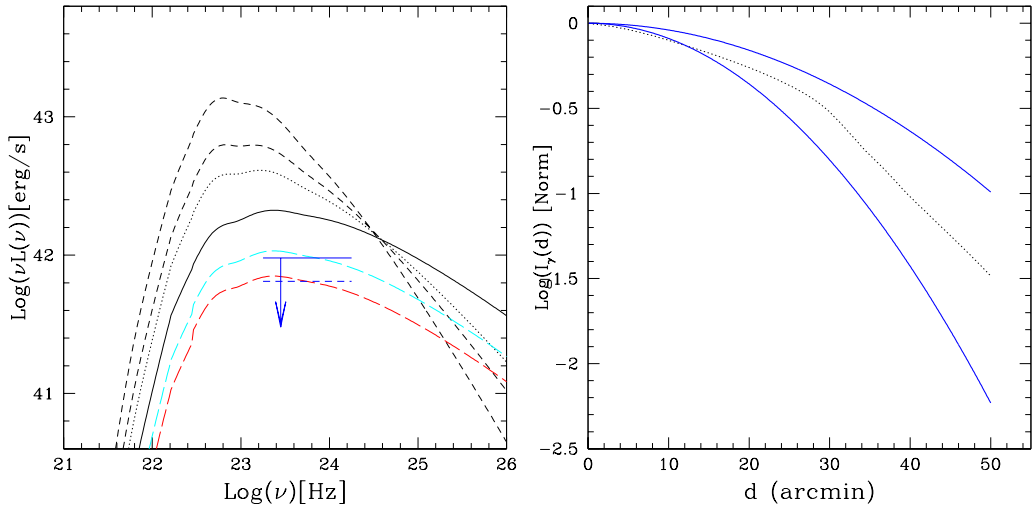}}
\end{center}
\caption{{\bf Left}: $\gamma$--ray spectrum due to $\pi^0$--decay
from hadronic models in Fig.~1.
The blue solid and dashed arrows mark the Ackermann et al.2010 
1--10 GeV limit and that extrapolated by assuming no detection 
after 3 years of FERMI observations, respectively.
To highlight the effect due to constraints from the radio brightness 
profile on the predicted $\gamma$--ray emission we also show
expectations for $\delta=2.6$ assuming the old brightness profile
from Deiss et al.~(1997) (cyan, long-dashed line) and assuming 
$f=0$ (red, long-dashed line).
{\bf Right}: The $\gamma$--ray brightness profile from hadronic models
in Fig.~1 compared with the Gaussians with 68\% surface
intensity containment radius $=$ 0.4 and 0.6 deg that are chosen 
in Ackermann et al. to derive the value of the Fermi-LAT upper limit
used in our paper.}
\label{fig:gamma}
\end{figure}

We adopt a general modeling of the non-thermal CR proton distribution 
in the Coma radio halo, $\epsilon_{CR} \propto \epsilon_{ICM}^{1+f}$ 
for $r \leq R_H$ ($R_H \sim 3 \, r_c$ being the halo radius) and 
$\epsilon_{CR} \propto \epsilon_{ICM}$ for $r > R_H$. 
This provides us with a conservative approach that minimizes the energy 
budget of CR protons in the very external regions ($r > 3-3.5 \, r_c$), 
where present data do 
not provide reliable constraints. Under these conservative assumptions the 
$\gamma$--ray brightness profile becomes steeper at projected distance 
$d \geq R_H$ where the brightness becomes proportional to that of the 
thermal X--ray emission; from Eq. \ref{jgamma} it is  
$I_{\gamma}(d)d^2 \propto d^{-6 \beta +3}$, implying that the 
majority of the $\gamma$--rays are produced within $\sim R_H$.

The observed radio luminosity at 350 MHz and the brightness profile 
allow us to constrain $K_p$ (see Sect.~2) and consequently to estimate 
the $\gamma$--ray luminosity for a given model of the cluster magnetic field. 
In order to break the degeneracy with the magnetic field properties, 
as a {\it reference point} we start by assuming the magnetic field strength 
and its spatial distribution as derived from the analysis 
of RM (Bonafede et al. 2010). Bonafede et al. analyzed polarization 
data for seven radio sources in the Coma cluster field observed with 
the VLA at 3.6, 6 and 20 cm. They derived Faraday rotation measures 
with kpc scale resolution and compared observations with simulations 
of random three-dimensional magnetic field models where the magnetic 
field was scaled with the thermal cluster density in the 
form $B(r) \propto B_0 n_{th}(r)^{\eta}$. 
They derive constraints for the magnetic field strength and 
profile $(B_0,\eta)$ with best values $B_0=4.7 \, \mu$G and $\eta=0.5$, 
respectively; $\eta \sim 0.5$ is also consistent with results 
from recent numerical MHD simulations of a variety of
clusters (Bonafede et al. 2011b).

\label{sec:hydro}
\begin{figure}
\begin{center}
{
\includegraphics[width=0.79\textwidth]{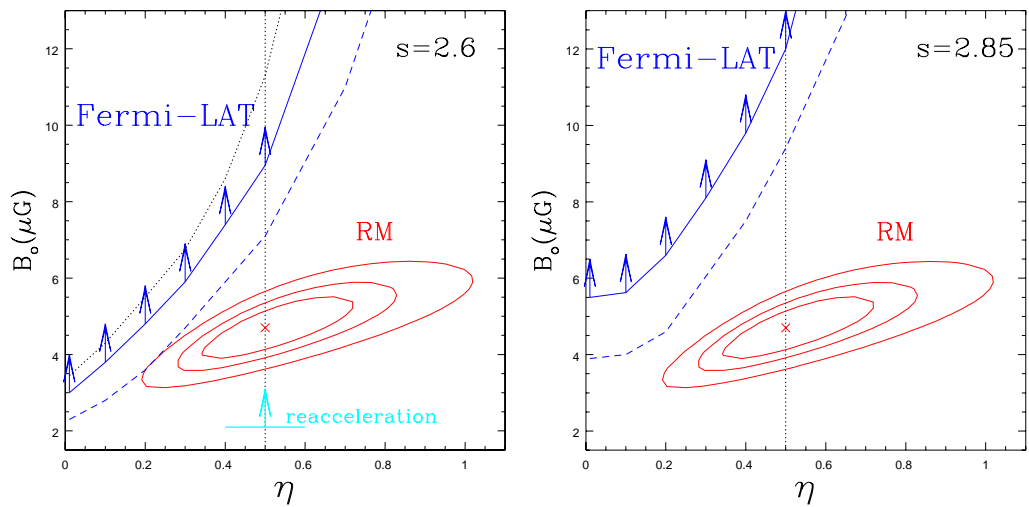}}
\end{center}
\caption{{\bf Left}: A comparison between the allowed 
region $(B_0, \eta)$ from the analysis of RM by Bonafede 
et al. 2010 (contours are reported at 1, 2, 3 $\sigma$) and 
that allowed, assuming a hadronic origin of the radio halo 
with a slope of CR protons 
$s=2.6$, by the 2--$\sigma$ upper limit from Ackermann 
et al. (dashed), assuming no detection from 3 years 
of Fermi-LAT observations (extrapolated from Ackermann, solid line),
and using the 2--$\sigma$ upper limit from Han et al.(2012) (dotted).
The vertical dotted line marks the relevant 
case $\eta =0.5$, in this case a lower limit to $B_0$ obtained 
by assuming a reacceleration model (Sect.3) is reported (cyan).
{\bf Right}: The same as in Left panel but assuming $s=2.85$; 
for simplicity here we do not report limits obtained by using 
Han et al. (2012).}
\label{fig:Bfield}
\end{figure}

The $\gamma$--ray emission from $\pi^0$ decay as calculated for 
the same choice of parameters as in Fig. \ref{fig:hadroComa} is 
shown in Fig. \ref{fig:gamma} (left panel) together with the Fermi-LAT 
upper limits from the first 18 month of observations 
(Ackermann et al. 2010). We are using the Ackermann et al. limits 
derived by chosing a Gaussian (with 68\% surface intensity containment radius
$=$ 0.4 and 0.6 deg) spatial distributions of $\gamma$-rays;
this is consistent with the spatial distribution expected from the model we 
adopted here (Fig. \ref{fig:gamma}, right panel). 
The comparison between the predictions of the model and the existing 
upper limits is therefore to be considered fair. 
In Fig. \ref{fig:gamma} (left panel, dashed) we also show the level of
the upper limit assuming no detection after 3 years of
observations;\footnote{Ando \& Nagai (2011)
and Han et al. (2012) presented a analysis of 3 years of Fermi-LAT
observations of nearby galaxy clusters.
They both report a 2$\sigma$ upper limit for the Coma cluster.}
that is obtained by simply scaling the Ackermann's upper limit
according to the square root of the exposure time.
For sake of clarity, 
in Fig.2 we also show the expected $\gamma$-rays in the case we assume
a constant ratio of the CR protons and thermal energy densities in the cluster
($f=0$, red long-dashed line), i.e. without considering the constraints 
from the radio brightness distribution of the Coma halo (see also
Sect.~4 and Fig.~4).

\noindent
Our results show that a pure hadronic origin of the halo is inconsistent 
with the Fermi-LAT upper limit, if the magnetic field properties are 
parametrized to satisfy the best fit obtained by Bonafede et al. (2010). 
Remarkably, the radio spectrum plotted in the left panel 
of Fig. \ref{fig:hadroComa} shows that the spectrum of the halo is 
best fitted assuming steeper CR spectra. 
This happens because the effect of the
negative flux from the thermal SZ-decrement becomes more relevant  
when the radio flux of the halo at high frequencies is smaller (see 
also Pfrommer \& En\ss lin 2004). However, steeper CR spectra produce 
more $\gamma$-rays at low energy and a hadronic origin of the halo 
is readily ruled out by using the Fermi-LAT upper limits.

The choice of the value of $\eta$ has a significant 
impact on the expected $\gamma$--ray emission from the cluster because
it determines the amount of CR protons that must be assumed 
at different distances from the cluster center to match the observed 
synchrotron profile.
There are some observational and theoretical arguments that suggest that
the ratio of turbulent and thermal pressure in clusters may increase
in the cluster external regions (e.g. Vazza et al 2009, Iapichino
et al 2011, Churazov et al 2012) which may eventually imply that the 
ratio of magnetic field and thermal energy densities 
could increase with cluster distance (assuming that magnetic field
and turbulent energy correlate).
For this reason we carry out more general calculations
assuming $\eta$ in the range 0--1.
This is shown in Fig. \ref{fig:Bfield} 
using present (Ackermann et al. 2010) Fermi-LAT upper limits (dashed line),
and assuming no detection of $\gamma$--rays from the Coma cluster 
after 3 years of observations (solid line, obtained by simply scaling 
the Ackermann et al. limit according to the square root of 
the exposure time); lower limits obtained by using very recent 
Fermi-LAT upper limits by Ando \& Nagai (2011) and Han et al. (2012) are 
also shown for the sake of completeness (dotted line).
For large values of $\eta$ ($\eta > 0.5$) the constraints are more stringent 
and only implausibly large values of $B_0$ are allowed. On the other hand 
for small values of $\eta$ fewer CR protons are required to match the radio 
luminosity and the brightness profile of the halo, leading to less stringent 
constraints on the magnetic field strength.
In particular the minimum value of $B_0$ increases with $\eta$,
from about $B_0 \geq 3 \mu$G for $\eta=0$ ($B_0 \geq 5.5 \mu$G for
$\delta =2.85$) to $B_0 \geq 8.5 \mu$G for $\eta=0.5$
($B_0 \geq 11 \mu$G for $\delta =2.85$).

\noindent
In Fig. \ref{fig:Bfield} we also show constraints on the
values of $B_0 - \eta$ as obtained by Bonafede et al.
We find disagreement between the region constrained from the
analysis of RM and that allowed by the lower limits on 
$B_0$ as obtained from the Fermi-LAT limits. 
In the relevant case 
$\epsilon_B \propto \epsilon_{ICM}$ ($\eta =0.5$), 
we find that the minimum
magnetic field energy density allowed by Fermi-LAT limits 
is 5-10 times larger than that of the {\it reference} magnetic field
energy density, estimated from the analysis of RM, while
for $\eta=0.2$ (still consistent with RM within 3$\sigma$) the
discrepancy is reduced to a factor $\sim 2$; the discrepancy
becomes severe for $\eta > 0.5$.
In general it is worth stressing here that our constraints using
upper limits from 3 years of Fermi-LAT observations select  
high magnetic fields, in which case even a moderate improvement
of the $\gamma$--ray limits implies a substantial increase of the 
minimum value of the magnetic field that is allowed (for
example, in the case $\eta \sim 0$ and $B_0 > B_{cmb}$
it is $B_{\rm min} \propto
{J_{\gamma}}_{(ul)}^{1/(1-\alpha) \sim -3}$, from Eqs.5--6).
Consequently deeper upper limits from Fermi-LAT in the next years,
or even a detection of the Coma cluster at a level 2-3 times below
Ackermann et al.~ limits will considerably increase the difficulties
in the context of a {\it pure} hadronic model for the radio halo.

\label{sec:hydro}
\begin{figure}
\begin{center}
{
\includegraphics[width=0.79\textwidth]{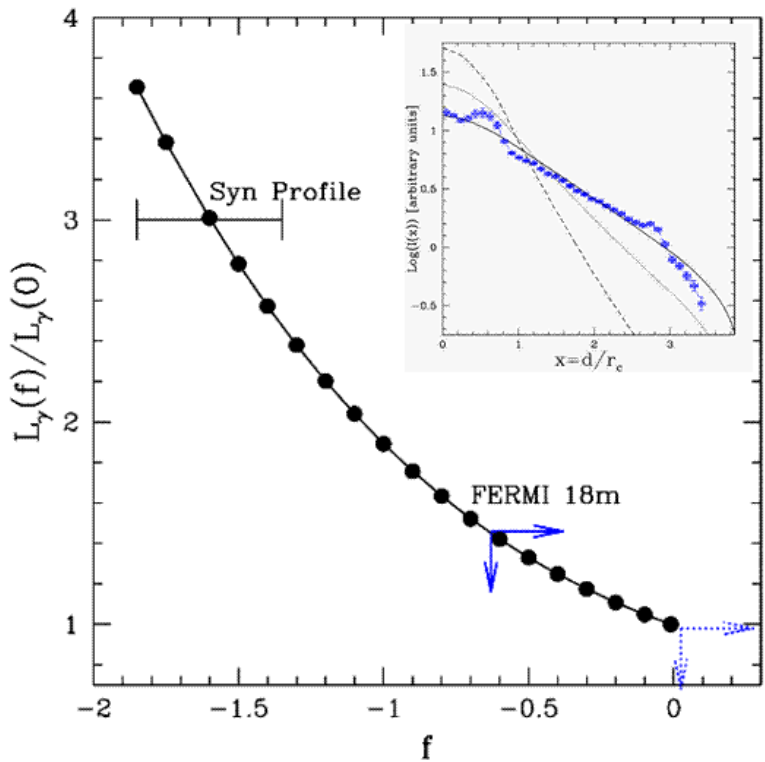}}
\end{center}
\caption{The $\gamma$-ray luminosity of the Coma cluster assuming hadronic 
models for the radio halo (assuming $B_0=4.7 \mu$G, $\eta=0.5$ and $s=2.6$) 
is shown as a function of $f$ and normalised to 
that expected for $f=0$ (namely assuming a scaling 
$\epsilon_{CR} \propto \epsilon_{ICM}$ as in 
Jeltema \& Profumo 2011). The inset shows the comparison between 
the observed brightness profile of the Coma halo and expectations based 
on hadronic models with $f=$-1.6, -1.0, 0 (from flatter to steeper). 
The horizontal error-bar gives the values of $f$ as constrained 
(3--$\sigma$) from the observed brightness profile (we consider the 
slope of the profile in the distance range 1--2.5 $r_c$). 
Blue arrows give the Fermi-LAT 
upper limits (solid$=$ Ackermann et al.(2010), 
dotted$=$ extrapolation of Ackermann's 
limit at 3 years observation) and their implications for the parameter $f$.}
\label{fig:gammaprofile}
\end{figure}

Using the Ackermann et al.~ upper limits, in the relevant case 
$\eta = 0.5$ Jeltema \& Profumo (2011) 
infer a value of the minimum magnetic field in the 
cluster center, $B_0 > 3.9 \mu$G, a result that is 
still consistent with the values 
constrained from RM. The difference between our work and 
Jeltema \& Profumo (2011) essentially stems from the adopted 
spatial distribution of CRs in the cluster. 
Jeltema \& Profumo (2011) assumed $\epsilon_{CRp} \propto \epsilon_{ICM}$  
while in our calculations we used the spatial distribution of CR protons 
needed to reproduce 
the observed synchrotron brightness profile of the radio 
halo. In order to highlight this point, in Fig. \ref{fig:gammaprofile} 
we show the ratio of the $\gamma$-ray luminosity that is obtained assuming 
a scaling $\epsilon_{CRp} \propto \epsilon_{ICM}^{1+f}$ and that assuming 
a linear scaling ($f=0$) between the energy densities of CRs and ICM. 
In Fig. \ref{fig:gammaprofile} we also report the constraints on $f$ 
derived from the synchrotron brightness profile of the halo and the limits 
implied by the Fermi-LAT upper limits after 18 months 
of observations (Ackermann et al. 2010) and by assuming no 
detection of the Coma cluster after $\sim 3$ years of observations.
We confirm that assuming $f=0$ (as in Jeltema \& Profumo) the value of 
the central magnetic field constrained from the Ackermann 
et al. $\gamma$-ray upper limits is still consistent with the value 
inferred from RM ($B_0 = 4.7 \mu$G, with $\eta =0.5$, 
Bonafede et al. 2010), however in this case the synchrotron brightness 
profile of the halo is predicted to be considerably steeper, 
inconsistent with the observed one (Fig. \ref{fig:gammaprofile}, inset).

Finally we stress here that 
the discussion above considers a {\it reference} value for 
the magnetic field in the Coma cluster based on an analysis of RMs.
These results are obtained by assuming that the RM originates entirely in 
the ICM (Bonafede et al. 2010).
If a contribution local to the radio galaxies is present, as found 
in other cases (e.g. 
Rudnick \& Blundell 2003, Guidetti et al. 2011), the values of the magnetic 
field in the ICM, as inferred from the observed RM, would be biased to higher
values (see Rudnick \& Blundell 2003) and the {\it reference} value 
should be considered as an upper limit to the actual cluster field.
This case would make our conclusions even stronger.

\section{Turbulent acceleration of secondary electrons 
and positrons}
\label{sec:reacc}

Models based on turbulent reacceleration 
of seed electrons (Schlickeiser et al. 1987,
Brunetti et al. 2001, Petrosian 2001, Fujita et al. 2003, Cassano
\& Brunetti 2005) rely upon the observed low efficiency of particle 
acceleration 
in the ICM, as suggested for instance by the 
$\geq 1$ GHz cutoff in the radio spectrum of 
Coma (Schlickeiser et al. 1987, Thierbach et al. 2003), and the fact
that radio halos are not common and are correlated with clusters' dynamics 
(see discussion in Brunetti et al. 2009). 
More recently, (indirect) 
evidence in favor of these models has been 
provided by the observation of radio halos with very steep 
spectra (Brunetti et al. 2008), interpreted as halos in which 
the radio spectrum starts steepening at frequencies smaller than 
the observing frequency, thereby favoring poorly efficient 
acceleration mechanisms.

Acceleration of electrons from the thermal pool to relativistic 
energies by MHD turbulence in the ICM faces serious 
energetic problems (e.g. Petrosian \& East 2008) and usually leads 
to large heating of the ICM (Blasi 2000, Petrosian \& East 2008). 
Consequently, turbulent acceleration models start from the assumption 
of a pre-existing population of relativistic particles that provides 
the seeds to ``reaccelerate'' during mergers. 
The combination of the ``unknown'' distribution of the seed particles 
in the ICM with the poorly known properties of turbulence in galaxy 
clusters reduces the predictive power of this scenario (see 
however Cassano et al.~2010b and references therein for predictions 
on the population properties of radio halos).

\noindent
In the context of these models an elegant possibility is based on 
clusters of galaxies as storage rooms of CRs  
(V\"olk et al. 1996, Berezinsky et al. 1997 
and Ensslin et al. 1997) and on the fact that 
the secondary electrons and positrons produced via inelastic collisions 
between CR protons and thermal protons in the ICM may be reaccelerated by
MHD turbulence during cluster mergers (Brunetti \& Blasi 2005, 
Brunetti \& Lazarian 2011). 
In this sense, this model is of a hybrid nature, but its requirements 
in terms of CR energy density in clusters are very meek. 
In this scenario, {\it off-state} (purely hadronic) radio halos in 
almost all galaxy clusters should have a radio luminosity 
$\sim$10 times smaller than that of classical {\it on-state} (turbulent) 
giant radio halos (Brunetti \& Lazarian 2011). 
The recent detection of diffuse emission in quiescent clusters from 
stacked analysis of the SUMSS survey may provide the first, possible 
support for this scenario (Brown et al. 2011).

\noindent
In these hybrid models $\gamma$-rays are produced as a result of 
the generation 
and decay of neutral pions and of IC emission from high 
energy ($\sim$ TeV) secondary electrons. This realization allows 
us to gather complementary tests for this scenario based on 
$\gamma$--ray observations. 
The level of $\gamma$-ray emission expected in this scenario is 
however significantly smaller than that in the pure hadronic case,
since the interaction between CR protons and their secondaries 
with the MHD turbulence enhances the ratio of radio (synchrotron) 
and $\gamma$-ray emission in connection with cluster 
mergers (Brunetti \& Lazarian 2011). Therefore it is easier to 
achieve consistency between the Fermi-LAT upper limits, radio observations 
of the spectrum, morphology and Faraday RM (see Brunetti 2011 for 
a recent review). 

In turbulent acceleration models the spectra and morphologies of 
radio halos depend on the combination of several physical quantities. 
To provide a simple estimate of 
the emissivity from a population 
of reaccelerated electrons we note that the turbulent 
energy flux that is damped by the coupling with relativistic 
electrons is eventually converted to synchrotron 
and IC radiation (Cassano et al.~2007): the bolometric synchrotron 
emissivity can be written as:

\begin{equation}
J_{syn} \propto {{F_{\epsilon_t} \Gamma_{e^{\pm}}/\Gamma_{th}}\over
{1 + (B_{cmb}/B)^2}},
\label{JsynT1}
\end{equation}
where $F_{\epsilon_t}$ is the energy injection rate of turbulence 
(per unit volume) and 
$\Gamma_{e^{\pm}}/\Gamma_{th} 
\propto (\epsilon_{e^{\pm}}/\epsilon_{ICM}) \sqrt{T}$ is the ratio of 
the damping rate of the turbulence by relativistic electrons 
and by thermal ICM (the second channel of 
damping goes into thermal heating). 
If the emitting electrons are secondary particles reaccelerated 
by the MHD turbulence and assuming that the 
reacceleration time, $\tau_{acc}$, is smaller than the cooling time, 
$\tau_{loss}$, of electrons with energy $E < E_{\nu}$ ($E_{\nu}$ being 
the energy of the electrons emitting at the observing frequency), 
we have $\epsilon_{e^{\pm}} \propto
\epsilon_{CRp} n_{th}$ and Eq. \ref{JsynT1} reads :

\begin{equation}
J_{syn}(\nu) \propto
n_{th}^{2+f} {{ B^2 }\over
{B^2 + B_{cmb}^2}} T^{3/2+f} \Theta (r,\alpha),
\label{JsynT2}
\end{equation}
where we made the simplified assumption 
$F_{\epsilon_t} \propto n_{th} T$ (see Cassano 
et al.~2007 for details). $\Theta$ is a fudge function to 
account for several effects. One effect is that in 
the internal, denser, region of the cluster Coulomb losses can play 
a role making the reacceleration of secondaries with initial energies 
$E \sim 100-300$ MeV more difficult. Another effect is that the damping 
rate of turbulence also depends on the ``spectral shape'' of 
relativistic (reaccelerated) particles (Brunetti \& Lazarian 2007) that 
is sensitive to the combination of physical parameters in the ICM. 
Based on previous calculations (Brunetti \& Blasi 2005, 
Brunetti \& Lazarian 2011) we expect that all these effects combine to
make the brightness profile flatter.

\noindent
Eq.~\ref{JsynT2} essentially implies that the brightness profile 
from reacceleration of secondary particles is very similar to that 
based on pure secondary  models (with $\alpha \sim 1$, Eq.~\ref{JsynS}), 
and that two regimes exist (considering $\Theta \sim$ constant, to
evaluate approximate scalings): 
(i) $I_{syn} \propto (1 + x^2)^{-3 \beta (2+f)/2 +1/2}$, 
for $B^2(x) \gg B_{cmb}^2$, and 
(ii) $I_{syn} \propto (1 + x^2)^{-3 \beta (2(1+\eta)+f)/2 +1/2}$, 
for $B^2(x) \ll B_{cmb}^2$.

Predictions from reacceleration of secondary particles differ radically
from those of pure 
hadronic models at high observing frequencies, where 
$\tau_{acc} \sim \tau_{loss}$ (or $\tau_{acc} > \tau_{loss}$). 
Under these conditions the synchrotron spectrum steepens at the 
frequency (Brunetti et al. 2001):

\begin{equation}
\nu_s \propto {{B \tau_{acc}^{-2}}\over{(B^2 + B_{cmb}^2)^2}},
\label{nus}
\end{equation}
In the IC-dominated regime, $B < B_{cmb}$, 
$\nu_s \propto B \tau_{acc}^{-2}$. Under the simple assumption that 
the magnetic field strength depends only on distance from the cluster 
center ($B(r)$, {\it homogeneus models}), the synchrotron spectrum steepens 
with distance (due to the fact that the emission is produced 
in a magnetic field that decreases with distance, and assuming an
approximately constant acceleration time--scale) and consequently radio 
halos become smaller at higher observing frequencies. 
We note indeed that observational claims exist for a decrease of 
the halo size with frequency and a steepening of the halo spectrum 
with projected distance (Giovannini et al. 1993, Deiss et al. 1997, see
also Fig.~1), 
although future observations are highly desirable to confirm these findings.

In this section we check that the predictions of the model based 
on turbulent reacceleration of secondary electrons is consistent 
with all existing observations and we investigate the implications 
of the model for the detectability of clusters in $\gamma$-rays. 
The turbulence and particle acceleration process are 
modelled following  
Brunetti \& Lazarian (2011)\footnote{We refer the readers to Brunetti \&
Lazarian 2011 for the formalism on turbulence and stochastic particle
re-acceleration. It allows us to model self-consistently the effect of
turbulent
re-acceleration on the spectra of CR protons, secondaries and 
of the non-thermal emission,
going beyond the simplified scalings in Eqs.\ref{JsynT1}--\ref{nus}}, 
where compressible turbulence is injected 
at large scales, $\sim$100-300 kpc, and decays onto smaller scales 
where it behaves as MHD turbulence. 
The cascading of fast modes is 
dissipated in the collisionless regime mainly due to 
transit--time--damping (TTD) with thermal and relativistic particles 
in the ICM. This process channels a fraction of the turbulent energy 
flux into the reacceleration of CR protons and secondary electrons. 
We carry out a homogeneous modeling of turbulence and turbulent 
reacceleration in the Coma halo, namely we assume a constant ratio 
$\epsilon_{tur}/\epsilon_{ICM}$, $\epsilon_{tur}$ being the turbulent 
energy density, and assume the same scalings between thermal and 
(the initial)
non-thermal properties as in Sect.~2.2. As a reference value we assume 
that turbulent motions on scales $l \leq 30$ kpc contribute to about 
5\% of the ICM energy (implying a total turbulent budget on the largest 
scales $\epsilon_{tur}/\epsilon_{ICM} \sim 0.2$)\footnote{This reference value
is needed
to reproduce radio halo properties, see Brunetti \& Lazarian 2007, 2011}.

\begin{figure}
\begin{center}
{
\includegraphics[width=0.79\textwidth]{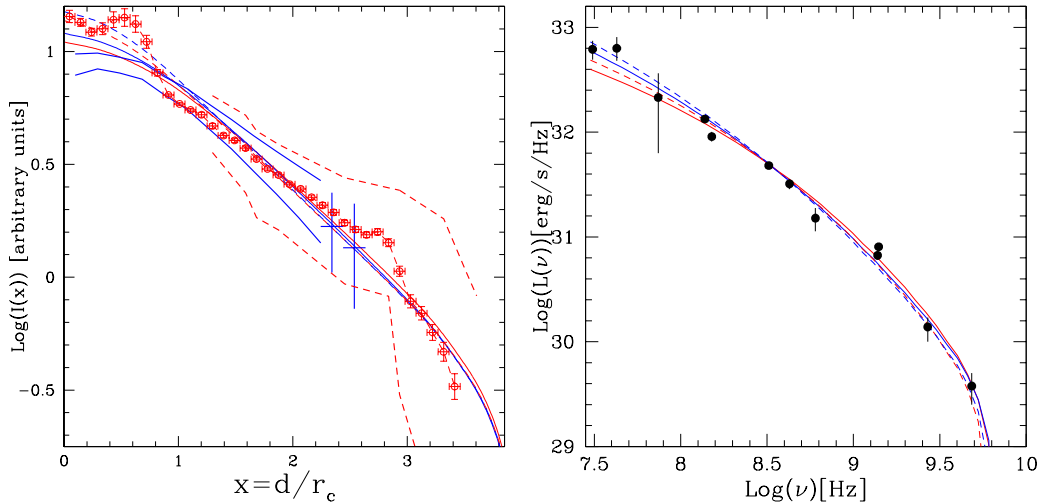}}
\end{center}
\caption{{\bf Left}: Brightness (azimuthal averaged) profile of the Coma
halo (as in Fig.~1) compared with expectations based on reacceleration models.
All models assume $\epsilon_{tur}/\epsilon_{ICM}\sim 0.05$ on scales 
$l < 30$ kpc and $s=$2.6. Calculations are carried out assuming
(i) $B_0 =5 \mu$G and $\eta=0.5$  (solid lines) with $f=-1.15$
and $\Delta \tau_r = 0.75$ Gyr (red lines) and $f=-1.3$ and
$\Delta \tau_r = 0.5$ Gyr (blue lines) ($\Delta \tau_r$ is the period
of reacceleration), and (ii) $B_0 =2 \mu$G and $\eta=0.5$ (dashed lines)
with $f=-1.6$ and $\Delta \tau_r = 0.75$ Gyr (red lines) and $f=-1.75$
and $\Delta \tau_r = 0.5$ Gyr (blue lines).
{\bf Right}: Spectrum of the Coma halo from reacceleration models as
in Left panel. All models are normalised at the radio halo luminosity
at 350 MHz.}
\label{fig:reacc}
\end{figure}

In Figure \ref{fig:reacc} we show the expected brightness profile 
(left panel) and synchrotron spectrum (right panel) of the Coma halo 
in the reference case $\eta =0.5$ and $B_0 \sim 5\mu$G. 
The spatial distribution of CR protons must be fairly flat, 
$f \sim -1.1$ to $-1.35$ (up to $r \sim 3 r_c$), so as to reproduce 
the observed radio profile.
This is similar to the case of pure hadronic models, although less
extreme because in the case 
of turbulent reacceleration the synchrotron profiles are slightly 
flatter (a spatially constant distribution of the energy density 
of CR protons, $f \sim -1$, is still consistent with observations) 
for the reasons discussed above
(see Eq.~\ref{JsynT2} and the paragraph below it).
Contrary to the case of pure
hadronic models, the steepening of the synchrotron 
spectrum at higher frequencies is reproduced very well
in the case of turbulent reacceleration 
models allowing a good representation of the data over almost 2 decades 
in frequency range; more (less) turbulence in 
the ICM would produce a steepening at higher (lower)
frequencies. 

\begin{figure}
\begin{center}
{
\includegraphics[width=0.79\textwidth]{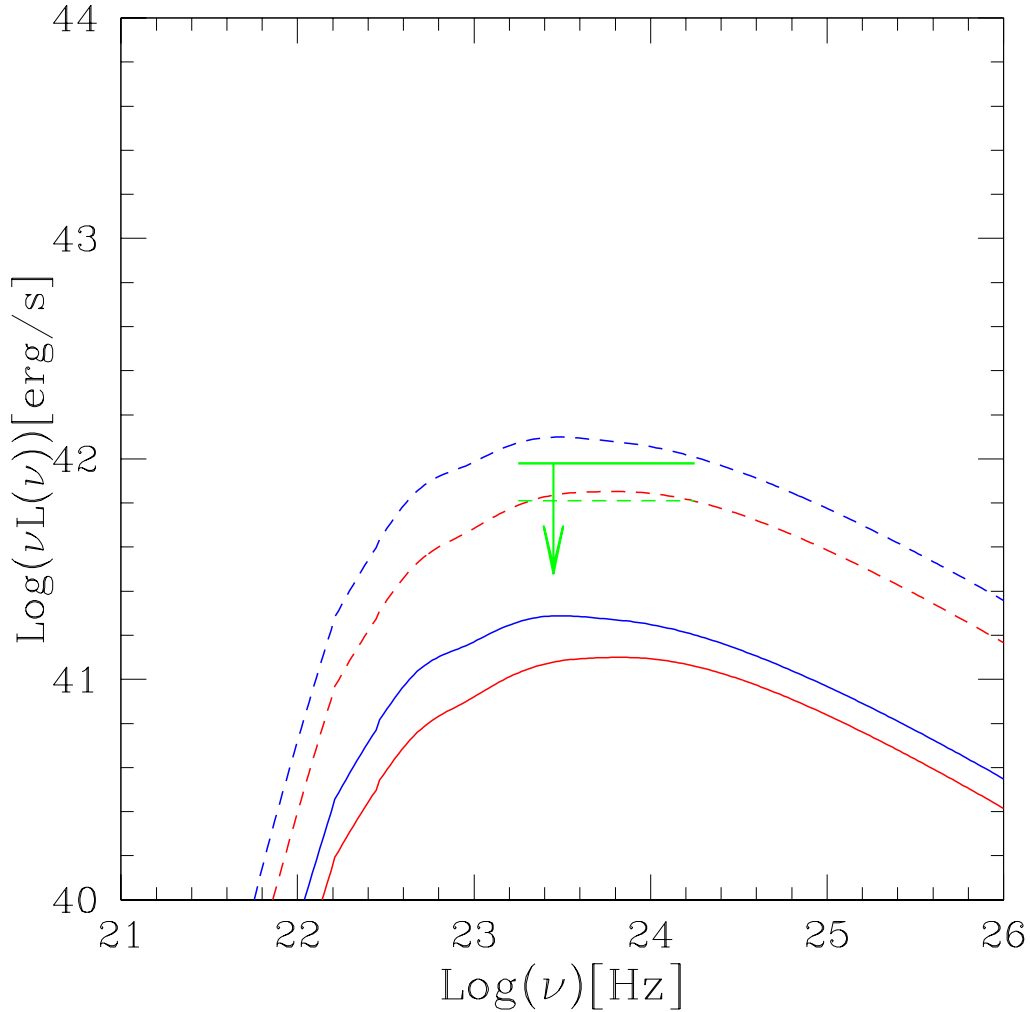}}
\end{center}
\caption{$\gamma$--ray spectrum due to $\pi^0$--decay as expected 
from reacceleration (hybrid) models. 
Lines are labelled as in Fig.~\ref{fig:reacc}.}
\label{fig:reaccgamma}
\end{figure}

During a re-acceleration phase 
the total energy budget of CR protons within the halo emitting 
volume ($r \sim 3 \,r_c$) is $\approx$4 \% of that of the ICM. 
This is almost 1 order of magnitude smaller than that in the case 
of pure hadronic models and implies a smaller $\gamma$--ray emission, 
well consistent with Fermi-LAT limits, as illustrated 
in Fig. \ref{fig:reaccgamma}.

More $\gamma$--rays are produced if the magnetic field in the halo 
volume is smaller. This allows us to constrain the minimum value
of the magnetic field that is required to have a $\gamma$--ray flux
consistent with Fermi-LAT limits, under the assumption that the halo
is generated by reacceleration of secondary particles.
In the relevant case $\eta=0.5$ we derive a minimum value 
$B_0 \geq 2 \mu$G (Fig.~ \ref{fig:reaccgamma}) that would imply a magnetic 
field energy density in the ICM $\sim 5$ times smaller than the best fit 
obtained from the analysis of Faraday RM by Bonafede et al.(2010).
For the sake of completeness, in Fig. \ref{fig:reacc} we show 
the expected synchrotron spectrum and the brightness profile of the 
Coma halo obtained by assuming $B_0 = 2 \mu$G and $\eta =0.5$. 
In this case the synchrotron emission is always produced under 
conditions $B_{IC}^2 \gg B^2$ and a significantly inverted 
spatial distribution of CR protons, $f \sim -1.5$ to $-1.8$
is needed to match the observed brightness profile, with the
consequence of a large energy budget in the form of CR protons.

\label{sec:hydro}

\section{Discussion and conclusions}
\label{sec:disc}

We presented a combined analysis of the spectrum and morphology of 
the giant radio halo in the Coma cluster, the available measurements 
of cluster magnetic fields from RM and the upper limits on the 
$\gamma$-ray emission from this cluster as recently obtained by the Fermi-LAT 
telescope (Ackermann et al. 2010)\footnote{see also 
Ando \& Nagai (2012) and Han et al. (2012)}, in the context of 
models based on secondary electrons: we concentrated on a pure secondary 
electron model and on a scenario where secondaries are reaccelerated by MHD
turbulence during mergers.

\subsection{Hadronic models}

\begin{figure}
\begin{center}
{
\includegraphics[width=0.79\textwidth]{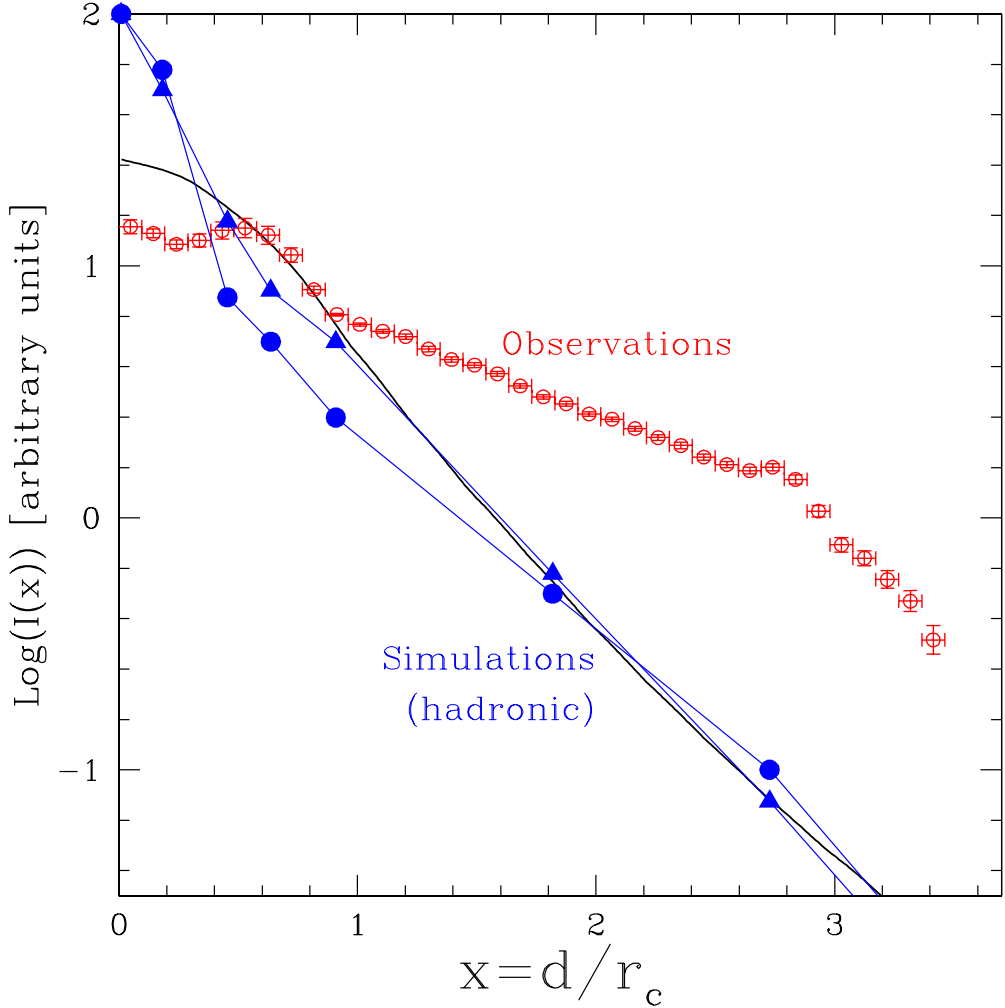}}
\end{center}
\caption{Azimuthal averaged brightness profile of the Coma halo 
(as in Fig.~1) compared 
with expectations based on numerical simulations that include the 
acceleration of CR protons and the generation of secondary electrons 
in the ICM. Points show the expected synchrotron profile from secondary 
electrons in the massive cluster gs72 from Pfrommer et al. 2008 (circles 
mark the case of radiative simulations). The solid line (black) show 
the expectations based on the semi-analytic
model of CR protons in Coma based on numerical 
simulations (Pinzke \& Pfrommer 2010).
A magnetic field profile $B(r)^2 \propto \epsilon_{ICM}$ and
$B_0 = 5 \mu$G are assumed.}
\label{fig:numerical} 
\end{figure}

The pure secondary electron model is based on the concept of CR effective 
confinement in the ICM (V\"olk et al. 1996, 
Berezinsky et al. 1997) and consists of explaining the giant radio halo 
emission as the result of synchrotron emission of secondary 
electrons (and positrons) from inelastic collisions of cosmic ray 
protons with gas in the ICM. These collisions result in the production 
and decay of charged and neutral pions. The former lead to production 
of secondary electrons, while the latter provide a channel of continuous 
production of gamma radiation. 
This model has been widely implemented also in cosmological simulations.
These simulations, that include, to some extent, CR physics
and the acceleration of CRs at cosmological shocks provided a picture
of the radio to $\gamma$--ray properties of galaxy clusters,
under the assumption that the emitting particles (electrons) are
generated through pp collisions in the ICM or accelerated
at shocks (e.g., Pfrommer et al. 2008 and references therein). 
The first simulations of this kind
predicted that clusters would be potentially
detectable in $\gamma$-rays with present
day $\gamma$--ray telescopes (Miniati 2003, Pfrommer 2008).
The most important assumption in these simulations is in the
efficiency of particle acceleration at weak shocks that is poorly known
(see Gabici \& Blasi 2003, 2004 and Kang et al.~2007 for a critical view).
More recently, numerical
simulations of large-scale structure formation made an attempt to reconcile
their results with the lack of detection of galaxy clusters
in the $\gamma$-ray band (Pinzke \& Pfrommer 2010,
see also Aleksic et al. 2010,11). Although these simulations provide
expectations (still) consistent with the Fermi-LAT upper limits,
it is worth mentioning that
they do not allow to reproduce the observed properties of radio halos if we
assume that halos originate via secondary emission,
in particular their (very broad) spatial extent (Donnert et al. 2010a,b;
Brown \& Rudnick 2011).
In this respect,
for the sake of completeness, in Fig. \ref{fig:numerical} we show a comparison
between the brightness profile of the Coma radio halo and expectations
based on CRs distributions from
Pfrommer et al. (2008) and from Pinzke \& Pfrommer (2010) simulations.
This can be made by using the semi-analytical prescription of
the spatial distribution of CR in galaxy clusters,
as derived from high resolution numerical simulations
of clusters (Pinzke \& Pfrommer 2010), and using parameters
of the Coma cluster to calculate the production rate of secondaries and the
synchrotron emissivity. 
According to these simulations the ratio of
the CR and thermal pressure in typical non-CC clusters is quasi constant
up to a distance $R/R_{vir} \sim 0.2$ and increases by a factor 2-3 up
to the virial radius (see Fig.14 in Pinzke \& Pfrommer 2010); such
an increase is however
mainly driven by the temperature decrement in the ICM at large distances.

\noindent
In reality, the microphysics of the ICM and of CRs in galaxy clusters
is very complicated and unfortunately
beyond the capabilities of present simulations.
For this reason in our paper we have carried out a analysis that
does not depend on the way CR protons are generated
in the ICM and on the complex processes of CR transport
and reacceleration that can take place in the cluster volume.
Rather than modeling these complex processes we indeed derived
constraints on the spectral and spatial distributions of CR protons 
directly from the observed spectrum and morphology of the Coma halo,
using parameters of the ICM from the X-ray observations (\S 2.2).

\noindent
We derived the azimuthally averaged brightness profile of the Coma halo
and used it in order to obtain constraints on the spatial profile of 
the magnetic field 
strength and of CRs density as functions of the distance from the 
cluster center.
Our profile is obtained using the deep WSRT observations from Brown \&
Rudnick (2011) and allows us to obtain unprecedented constraints
on the non-thermal cluster properties on 3--3.5 $r_c$ scales.
We find that fitting the volume averaged radio spectrum of the Coma halo 
and its azimuthal brightness distribution requires a rather steep spectrum 
of CR protons, $s\sim 2.6$, in the ICM with a 
spatial distribution much flatter than the spatial distribution 
of the thermal gas. This leads 
to an exceedingly large energy content in the form of CRs confined 
in the ICM and correspondingly large $\gamma$-ray emission, 
exceeding the Fermi-LAT upper limits. Moreover the spectral steepening 
observed at high frequency can only marginally be explained for 
the cases with steeper slope ($s \geq 3$) and the combined effect of 
the SZ decrement. This case is however the one that violates the 
Fermi-LAT upper limits more clearly. 

\noindent
If we neglect the spectral steepening in our analysis, the model
can potentially explain radio observations. However the Fermi-LAT 
$\gamma$--ray limits set a lower limit to the strength of the
cluster magnetic field that is in disagreement with a recent 
analysis of Faraday
rotation measures (Bonafede et al. 2010);
the more so for steeper spectra of the parent CR protons.

\noindent
{\it Pure} secondary electron models of the Coma halo
are thus disfavoured when all available observational data
are considered.
In this respect any further improvement of upper limits over the next
several years, or even a $\gamma$-ray detection of the Coma cluster 
(i.e. with flux 2-3 times smaller the Ackermann et al. limits)
will conclusively establish the incompatibility of a hadronic 
origin of the Coma radio halo with the radio (including RM) and $\gamma$-ray 
observations.

\subsection{A comment on the most relevant assumptions}

\noindent
A discussion of the assumptions adopted in our analysis is in order. 
Conclusions derived above are based on canonical assumptions used to
model non-thermal emission in galaxy clusters.
The most notable assumption is that the CR spectrum in the cluster volume
is a power law in momentum, 
$N(p) \propto p^{-s}$, and that the spectrum of secondary particles
emitting in the radio band can be calculated under stationary
conditions. We constrained the value of the slope $s$ from 
the spectrum of the radio halo in the frequency 
range $\sim$0.1-1 GHz (Fig. \ref{fig:hadroComa}) and the $\gamma$-ray 
emission of the cluster is calculated by using such values of $s$. 
In principle, the shape of the spectrum of CR protons 
might also change with location,
moving away from cluster core toward the periphery, although there is 
no strong 
indication that this may be happening from the total radio spectrum of the
Coma halo\footnote{Indeed a mixture of constributions from different power-law 
spectra results in a ``concave'' shape of the total synchrotron 
spectrum.}.
In this case the expected $\gamma$--ray emission is reduced
if the spectrum of CR protons is flatter in the regions where the majority of 
$\gamma$--rays are produced, at distances 2--3.5 $r_c$ from the center. 
This however would result in a halo spectrum that is flatter in the
external regions, in contrast with present 
observations that suggest a ``radial spectral steepening''
(Giovannini et al 1993, Deiss et al 1997, this is also evident from
the comparison between the synchrotron brightness distributions of
the halo at 350 and 1400 MHz, Fig. 1).

\noindent
The situation may change if the spectrum of CRs becomes harder at low enough 
energies. In this case one may expect that the flux of low energy 
$\gamma$-rays 
(from decays of neutral pions) is also reduced allowing for some room
for pure hadronic models to accomodate current observational constraints.
Present radio data limit the 
possibility of a flattening in the radio spectrum to frequencies 
$\nu \leq 100$ MHz, leading to a limit to the energy where a possible 
break occurs in the spectrum of the primary CR protons, 
$E_b \leq 10-20$ GeV. Under these conditions we note that
a substantial reduction of the $\gamma$-ray luminosity due to $\pi^0$--decay 
in the 1-10 GeV band, necessary to make hadronic models still consistent 
with --at least-- the limits derived from the first 18 month of FERMI-LAT data
(Ackermann et al. 2010) (Fig. \ref{fig:gamma}), would require 
a prominent CR spectral break, by $\Delta s \ge 0.5$.

The two points that disfavour pure hadronic models for the radio halo
are the steepening of the halo spectrum observed at higher frequencies and 
the inconsistency between the properties of the cluster magnetic field 
constrained by Faraday RM and by Fermi-LAT limits under the assumption 
of a hadronic origin of the halo.
The latter point however might simply indicate that RM and synchrotron
emission trace different magnetic fields. 
As already mentioned in Sect. 2.3 the simplest way to have RM and 
synchrotron emission sample different fields is to assume that some of 
the RM come from regions adjacent to and influenced by the radio sources, 
in this case however the magnetic field in the ICM as inferred from the 
analysis of RM would likely be biased to higher values of the magnetic 
field (Rudnick \& Blundell 2003) thus making the inconsistency between 
magnetic fields even stronger. 
On the other hand, 
if one assumes that RM originates entirely in the ICM, RM and synchrotron 
emission may sample different fields in the case of a highly inhomogeneous 
fields, because the synchrotron emissivity depends non-linearly on the 
magnetic field and its fluctuations.
Positive fluctuations however increase also the radiative losses of particles
(assuming they vary on time scale sufficiently long,
$> 10^7-10^8$yrs\footnote{The minimum RM scale of about 2 kpc for Coma
radio galaxies derived by Bonafede et al 2010, and the typical fractional
polarization of about 10\%, imply that any ICM fluctuation on smaller
scales, that could be very intermittent, must be weak.}) and this is
expected to partially quench the expected boosting of the synchrotron
emissivity of electrons generated in regions with medium/high fields.
Under (at least quasi--) stationary conditions the emissivity reads:

\begin{equation}
<J_{syn}>_{Volume} = \propto
{\hat B}^{1+\alpha} {{ ( 1 + (\delta B/{\hat B})^2)^{(1+\alpha)/2} }
\over{ {\hat B}^2 + \delta B^2 + B^2_{cmb} }} \, ,
\label{jsyn2}
\end{equation}

\noindent
where we introduced a magnetic field made of a large scale 
component ${\hat B}$ and a turbulent component 
$\delta B$, such that $<\bf{ {\hat B}} + \bf{\delta B}>_{Volume}=
{\bf {\hat B}}$ and 
$<\bf{ {\hat B}} + \bf{\delta B}>^2_{Volume}= {\hat B}^2 + \delta B^2$.
Under these assumptions we estimate that the ratio of $\gamma$--ray and 
radio cluster luminosities decreases by (only) a factor 
$\sim 1.7$, compared to results using the formalism in Sect.~2.1, if we 
assume $\delta B^2 \sim {\hat B}^2$, 
with $\delta B^2$ anchored to the best fit value from RM studies
(normalisation and spatial profile). 

On the other hand, RMs 
suggest that $\delta B\gg B$, in which case Eq.\ref{jsyn2} becomes
equivalent to Eq.\ref{jsyn} in Sect.~2.1\footnote{
In the analysis by Bonafede et al. the magnetic field is 
parameterised with a power spectrum $B^2 = 8\pi \int P_B(k) dk$
between a maximum and minimum scale and its properties are constrained 
by comparing 
observations with simulated Faraday Rotation maps derived for different 
parameters}.
In this sense the question of whether Faraday RM and synchrotron emission 
measure the same magnetic field only lies in the capability of RM studies 
to provide a good description of the ICM magnetic field.
Future radio telescopes, including LOFAR and SKA, will greatly improve the
sensitivity to RMs allowing to detect and use many tens of background 
radio sources per clusters to sample magnetic fields along 
many lines of sight, thus removing potential biases in present studies.

\noindent
Finally, for the sake of completeness we mention that in principle, spatial 
diffusion of particles in inhomogeneous fields may also affect the value 
of the ratio synchrotron/$\gamma$-ray luminosity in hadronic models, 
if the diffusion 
time necessary to cover the spatial scales on which field inhomogeneities 
occur is smaller 
than both the life-time of particles and the time-scale of magnetic 
field (local) variations. This however depends on the details of the 
magnetic field and diffusion model.

\subsection{Beyond the pure hadronic model: turbulent reacceleration 
of secondaries}

In \S \ref{sec:reacc} we went beyond the pure hadronic model and
discussed the case of the reacceleration model were secondary particles
are reaccelerated by MHD turbulence.
We find that this model allows to obtain a good
description of the radio spectrum and morphology of the Coma 
cluster and at the same time they may easily be compatible with 
the Fermi-LAT upper limits on the $\gamma$-ray emission from this cluster 
and with the measured magnetic field strength as inferred from RMs. 

\noindent
In the simplified assumption that both the ratios of turbulent and thermal 
energy density, $\epsilon_{tur}/\epsilon_{ICM}$, and of magnetic field
and thermal energy density , $\epsilon_{B}/\epsilon_{ICM}$, are 
constant in the radio emitting volume, the brightness profile of the 
Coma halo leads us to infer 
that the spatial distribution of CRs in the cluster should be very broad, 
flat (or slightly increasing in the external regions) on the halo size-scale.
This is because the particular reacceleration model adopted in this 
paper faces drawbacks that are in part similar to those of a pure hadronic
model for the origin of the halo. 
This is simply because the seed electrons 
for reacceleration are generated by proton-proton collisions in the ICM.
A flat spatial distribution of CRs is a fairly strong requirement, although
possible support for a rather flat distribution of CRs,
significantly broader than that of the thermal ICM,
comes from very recent numerical cosmological simulations
that include CRs accelerated at shocks (Vazza et al.~2012).
Present radio data do not constrain the energy density of CR protons
on scales larger than the halo size-scale. However if we speculate that 
the flat spatial distribution of CR protons extends on larger scales the 
resulting energy budget of CRs in the cluster would be significantly
larger than that required by our modeling.

In general several, effects may contribute to mitigate this 
situation. 
First, numerical simulations show that the turbulent 
energy density (and the ratio $\epsilon_{tur}/\epsilon_{ICM}$) increases 
outside the cores of simulated clusters (Vazza et al. 2009, 11; 
Iapichino \& Niemeyer 2008, Iapichino et al. 2011), implying a synchrotron 
brightness profile potentially flatter than that calculated under 
the assumption of a constant ratio $\epsilon_{tur}/\epsilon_{ICM}$. 
Second, we limit our analysis to the case $B^2 \propto \epsilon_{ICM}$
($\eta =0.5$), that gives the best fit to Faraday RM. 
On the other hand for smaller values of $\eta$ the spatial distribution 
of CR protons that is required by the model to match the observed synchrotron 
brightness profile of the halo is significantly steeper with radius, 
although still flatter than the distribution of the thermal energy density
of the cluster.
Finally we note that the situation is greatly mitigated as soon as
one would relax the assumption of having only secondary electrons
in the ICM.
Indeed primary electrons accelerated at shocks, or during the 
activity of cluster galaxies and AGN, can survive a substantial fraction 
of the Hubble time in the cluster outskirts (e.g. Sarazin 1999). 
These primaries provide a natural population of seed electrons to 
reaccelerate in a turbulent ICM (Brunetti et al 2001, Petrosian 2001)
and may significantly contribute to 
the synchroton emission in the external regions of giant radio halos. 
If a substantial contribution to the radio halo emission comes from 
the reacceleration of primary electrons, the expected $\gamma$--ray 
emission from the Coma cluster becomes even smaller than that 
calculated in \S \ref{sec:reacc}. 

\noindent
Ongoing and future observations in the deep non-thermal (high and very-high
energy) regime of the Coma cluster will therefore provide 
precious information on its non-thermal content.

\section*{Acknowledgements}
Authors acknowledge useful comments from the anonymous referee and 
discussions with J. Donnert and T. Jeltema.
GB acknowledges partial supported by INAF under grant PRIN-INAF 2009.
LR acknowledges support, in part, by U.S. National Science Foundation grant
AST-0908688 to the University of Minnesota.
AB acknowledges partial supported by the DFG Research Unit 1254
``Magnetization of interstellar and intergalactic media: the 
prospect of low frequency radio observations''.


\begin{thebibliography}{}
\bibitem{} Ackermann M., et al. 2010, ApJ 717, L71
\bibitem{} Ando, S., \& Nagai, D.\ 2012, arXiv:1201.0753 
\bibitem{} Aleksic J., et al. 2010, ApJ 710, 634
\bibitem{} Aleksi{\'c}, J., Alvarez, E.~A., et al.\ 2011, arXiv:1111.5544 
\bibitem{} Aharonian F.A., et al., 2009a, A\&A 495, 27
\bibitem{} Aharonian F.A., et al., 2009b, A\&A 502, 437
\bibitem{} Berezinsky V.S., Blasi P., Ptuskin V.S., 1997, ApJ 487, 529
\bibitem{} Blasi, P.\ 2000, ApJ, 532, L9 
\bibitem{} Blasi P., 2001, APh 15, 223
\bibitem{} Blasi P., Colafrancesco S., 1999, APh 12, 169
\bibitem{} Blasi P., Gabici S., Brunetti G., 2007, IJMPA 22, 681
\bibitem{} Bonafede A., Feretti L., Murgia M., Govoni F., Giovannini G.,
Dallacasa D., Dolag K., Taylor G.B., 2010, A\&A 513, 30
\bibitem{} Bonafede, A., Govoni, F., Feretti, L.,
et al.\ 2011a, A\&A, 530, A24 
\bibitem{} Bonafede A., Dolag K., Stasyszyn F., Murante G.,
Borgani S., 2011b, MNRAS 418, 2234
\bibitem{} Briel U.G., Henry J.P., B\"ohringer H., 1992, A\&A 259, L31
\bibitem{} Brown S., Rudnick L., 2011, MNRAS, 412, 2
\bibitem{} Brown, S., Emerick, A., Rudnick, L., \& Brunetti, G.\ 2011, 
ApJ, 740, L28 
\bibitem{} Brunetti, G.\ 2004, JKAS, 37, 493 
\bibitem{} Brunetti G., 2009, A\&A 508, 599
\bibitem{} Brunetti, G.\ 2011, MmSAI, 82, 515 
\bibitem{} Brunetti G., Setti G., Feretti L., Giovannini G., 2001, MNRAS
320, 365
\bibitem{} Brunetti G., Blasi P., 2005, MNRAS 363, 1173
\bibitem{} Brunetti G., Lazarian A., 2007, MNRAS, 378, 245
\bibitem{} Brunetti, G., Venturi, 
T., Dallacasa, D., et al.\ 2007, ApJ, 670, L5 
\bibitem{} Brunetti G., Giacintucci S., Cassano R., Lane W., Dallacasa
D., Venturi T., Kassim N.E., Setti G., Cotton W.D., Markevitch M., 2008,
Nature 455, 944
\bibitem{} Brunetti G., Cassano R., Dolag K., Setti G., 2009, A\&A 507, 661
\bibitem{} Brunetti G., Lazarian A., 2011, MNRAS, 410, 127
\bibitem{} Cassano R., Brunetti G., 2005, MNRAS 357, 1313
\bibitem{} Cassano, R., Brunetti, 
G., Setti, G., Govoni, F., \& Dolag, K.\ 2007, MNRAS, 378, 1565 
\bibitem{} Cassano, R., et al., 2008, A\&A, 480, 687
\bibitem{} Cassano R., Ettori S., Giacintucci S., Brunetti G.,
Markevitch M., Venturi T., Gitti M., 2010a, ApJ 721, L82
\bibitem{} Cassano, R., et al., 2010b, A\&A, 509, 68
\bibitem{} Churazov E., Vikhlinin A., Zhuravleva I., et al., 2012, MNRAS
421, 1123
\bibitem{} Colafrancesco, S., \& Blasi, P.\ 
1998, Astroparticle Physics, 9, 227 
\bibitem{} Colafrancesco, S., Marchegiani, P., Perola, C., 2005, A\&A 443, 1
\bibitem{} Dallacasa D., et al., 2009, ApJ, 699, 1288
\bibitem{} Deiss, B.~M., Reich, W., Lesch,
H., \& Wielebinski, R.\ 1997, A\&A, 321, 55 
\bibitem{} Dennison B., 1980, ApJ, 239, L93
\bibitem{} Dermer C.D., 1986a, ApJ 307, 47
\bibitem{} Dermer C.D., 1986b, A\&A 157, 223
\bibitem{} Dolag K., Ensslin T.A., 2000, A\&A 362, 151
\bibitem{} Donnert J., Dolag K., Brunetti G., Cassano R., Bonafede A.,
2010a, MNRAS 401, 47
\bibitem{} Donnert J., Dolag K., Cassano R., Brunetti G., 2010b, MNRAS
407, 1565
\bibitem{} En{\ss}lin, T.~A. 2002, A\&A, 396, L17 
\bibitem{} En{\ss}lin, T.~A., 
Biermann, P.~L., Kronberg, P.~P., \& Wu, X.-P.\ 1997, ApJ, 477, 560
\bibitem{} En{\ss}lin, T., Pfrommer, C., Miniati,
F., \& Subramanian, K.\ 2011, A\&A, 527, A99 
\bibitem{} Feretti, L., Fusco-Femiano, R.,
Giovannini, G., \& Govoni, F.\ 2001, A\&A, 373, 106 
\bibitem{} Ferrari, C.; Govoni, F.; Schindler, S.; Bykov, A. M.; Rephaeli,
Y., 2008, SSRv 134, 93
\bibitem{} Fujita Y., Takizawa M., Sarazin C.L., 2003, ApJ 584, 190
\bibitem{} Gabici S., Blasi P., 2003, ApJ 583, 695
\bibitem{} Gabici, S., \& Blasi, P.\ 2004,
Astroparticle Physics, 20, 579 
\bibitem{} Giovannini G., Feretti L., Venturi T., Kim K.-T., Kronberg P.P.,
1993, ApJ 406, 399
\bibitem{} Giovannini, G., Tordi, M., Feretti, L., 1999, NewA, 4, 141
\bibitem{} Giovannini, G., et al., 2009, A\&A, 507, 1257
\bibitem{} Govoni F., Ensslin T.A., Feretti L., Giovannini G., 2001,
A\&A 369, 441
\bibitem{} Guidetti, D., Laing, 
R.~A., Bridle, A.~H., Parma, P., \& Gregorini, L.\ 2011, MNRAS, 413, 2525 
\bibitem{} Han, J., Frenk, C.~S., Eke, V.~R., Gao, L., \& 
White, S.~D.~M.\ 2012, arXiv:1201.1003 
\bibitem{} Jeltema T.E., Profumo S., 2011, ApJ, 728, 53
\bibitem{} Kang, H., Ryu, D., Cen, R., \& Ostriker, J.~P.\ 2007, ApJ, 
669, 729 
\bibitem{} Kempner, J.~C., Sarazin, C.~L., 2001, ApJ, 548, 639
\bibitem{} Keshet U., 2010, arXiv:1011.0729 
\bibitem{} Keshet U., Loeb A., 2010, ApJ, 722, 737
\bibitem{} Kushnir, D., Katz, B., \& Waxman, E.\ 2009, JCAP, 9, 24 
\bibitem{} Iapichino, L., Niemeyer, J.~C., 2008, MNRAS, 388, 1089
\bibitem{} Iapichino, L., Schmidt, W., Niemeyer, J.~C., \& 
Merklein, J.\ 2011, MNRAS, 414, 2297 
\bibitem{} Macario G., Venturi T., Brunetti G., Dallacasa D., Giacintucci S.,
Cassano R., Bardelli S., Athreya R., 2010, A\&A, 517, 43
\bibitem{} Marchegiani P., Perola G. C., Colafrancesco S., 2007, A\&A 465, 41
\bibitem{} Miniati, F.\ 2003, MNRAS, 342, 1009 
\bibitem{} Moskalenko I.V., Strong A.W., 1998, ApJ 493, 694
\bibitem{} Murgia, M., et al., 2009, A\&A, 499, 679
\bibitem{} Perkins, J.~S., Badran, 
H.~M., Blaylock, G., et al.\ 2006, ApJ, 644, 148 
\bibitem{} Petrosian V., 2001, ApJ 557, 560
\bibitem{} Petrosian V., East W.E., 2008, ApJ 682, 175
\bibitem{} Pfrommer C., 2008, MNRAS 385, 1242
\bibitem{} Pfrommer C., Ensslin T.A., 2004, MNRAS 352, 76
\bibitem{} Pfrommer C., Ensslin T.A., Springel V., 2008, MNRAS 385, 1211
\bibitem{} Pinzke, A., \& Pfrommer, C.\ 2010, MNRAS, 409, 449 
\bibitem{} Pizzo, R., 2010, {\it Tomography of galaxy clusters through
low-frequency radio polarimetry}, PhD Thesis, Groningen University
\bibitem{} Reimer A., Reimer O., Schlickeiser R., Iyudin A., 2004,
A\&A 424, 773
\bibitem{} Reimer O., Pohl M., Sreekumar P., Mattox J.R., 2003, ApJ 588, 155
\bibitem{} Ribicky G.B., Lightmann A.P., {\it Radiative Processes
in Astrophysics}, New York, Wiley-Interscience, 1979
\bibitem{} Rudnick, L., \& Blundell, K.~M.\ 2003, ApJ, 588, 143 
\bibitem{} Sarazin, C.L. 1999, ApJ 520, 529
\bibitem{} Sarazin, C.~L.\ 2004, JKAP, 37, 433 
\bibitem{} Schlickeiser R., Sievers A., Thiemann H.: 1987, A\&A 182, 21
\bibitem{} Thierbach M., Klein U., Wielebinski R., 2003, A\&A 397, 53
\bibitem{} van Weeren, R.~J., Br{\"u}ggen, M.,
R{\"o}ttgering, H.~J.~A., et al.\ 2011, A\&A, 533, A35
\bibitem{} Vazza, F., Brunetti, G., Kritsuk, A., et al.\ 2009, A\&A, 504, 33 
\bibitem{} Vazza, F., et al., 2011, A\&A, 529, 17
\bibitem{} Vazza, F., Bruggen, M., Gheller, C., \& Brunetti, G.\ 2012, 
arXiv:1201.3362 
\bibitem{} Venturi, T.\ 2011, MmSAI, 82, 499 
\bibitem{} V\"{o}lk H.J., Aharonian F.A., Breitschwerdt D., 1996, SSRv 75, 279
\bibitem{} V{\"o}lk, H.~J., \& Atoyan, A.~M.\ 1999, Astroparticle 
Physics, 11, 73 
\bibitem{} Willson, M.~A.~G., 1970, MNRAS, 151, 1
\bibitem{} Wolfe, B., Melia, F., 
Crocker, R.~M., \& Volkas, R.~R.\ 2008, ApJ, 687, 193 
\end{thebibliography}
\end{document}